\documentclass[prb,superscriptaddress,twocolumn]{revtex4}
\usepackage{bm,graphicx,graphics,amsmath,amssymb,bm,epsfig,color}
\usepackage{euscript,tabularx}
\usepackage{textcomp}
\usepackage{gensymb}

\begin{document}
\bibliographystyle{apsrev}

\preprint{\begin{minipage}{8cm} Number: \hrulefill 
   \end{minipage}}

 \title{Ferromagnetic resonance force spectroscopy of individual
   sub-micron size samples}

\author{O. Klein}
\thanks{Corresponding author} \email{oklein@cea.fr}
\affiliation{Service de Physique de l'{\'E}tat Condens{\'e} (CNRS URA 2464), CEA Saclay, 91191 Gif-sur-Yvette, France}

\author{G. de Loubens}
\affiliation{Service de Physique de l'{\'E}tat Condens{\'e} (CNRS URA 2464), CEA Saclay, 91191 Gif-sur-Yvette, France}

\author{V. V. Naletov}
\affiliation{Service de Physique de l'{\'E}tat Condens{\'e} (CNRS URA 2464), CEA Saclay, 91191 Gif-sur-Yvette, France}
\affiliation{Physics Department, Kazan State University, Kazan 420008,
  Russia}

\author{F. Boust}
\affiliation{ONERA, 29 avenue de la Division Leclerc, 92322 Ch\^atillon, France}

\author{T. Guillet}
\affiliation{Service de Physique de l'{\'E}tat Condens{\'e} (CNRS URA 2464), CEA Saclay, 91191 Gif-sur-Yvette, France}

\author{H. Hurdequint}
\affiliation{Laboratoire de Physique des Solides, Universit\'e Paris-Sud, 91405 Orsay, France}

\author{A. Leksikov}
\affiliation{Service de Physique de l'{\'E}tat Condens{\'e} (CNRS URA 2464), CEA Saclay, 91191 Gif-sur-Yvette, France}

\author{A. N. Slavin}
\affiliation{Department of Physics, Oakland University, Michigan 48309, USA}

\author{V. S. Tiberkevich}
\affiliation{Department of Physics, Oakland University, Michigan 48309, USA}

\author{N. Vukadinovic}
\affiliation{Dassault Aviation, 78 quai Marcel Dassault, 92552 Saint-Cloud, France}

\date{\today}

\begin{abstract}
  We review how a magnetic resonance force microscope (MRFM) can be
  applied to perform ferromagnetic resonance (FMR) spectroscopy of
  \emph{individual} sub-micron size samples. We restrict our attention
  to a thorough study of the spin-wave eigen-modes excited in
  permalloy (Py) disks patterned out of the same 43.3 nm thin film.
  The disks have a diameter of either $1.0$ or $0.5~\mu$m and are
  quasi-saturated by a perpendicularly applied magnetic field. It is
  shown that \emph{quantitative} spectroscopic information can be
  extracted from the MRFM measurements. In particular, the data are
  extensively compared with complementary approximate models of the
  dynamical susceptibility: i) a 2D analytical model, which assumes an
  homogeneous magnetization dynamics along the thickness and ii) a
  full 3D micromagnetic simulation, which assumes an homogeneous
  magnetization dynamics below a characteristic length scale $c$ and
  which approximates the cylindrical sample volume by a discretized
  representation with regular cubic mesh of lateral size $c=3.9$ nm.
  In our analysis, the distortions due to a breaking of the axial
  symmetry are taken into account, both models incorporating the
  possibility of a small misalignment between the applied field and
  the normal of the disks.
\end{abstract}

\maketitle

\section{Introduction}

Development of innovative tools capable of measuring the local
magnetization dynamics $\bm{M} (t, \bm r)$ inside a ferromagnetic
nano-structure is an objective of primary importance. New technology
fields related to magnetic materials, like spintronics, depend on
one's ability to analyze and predict the out-of-equilibrium state of
$\bm{M}$ in nanoscale hybrid structures.  In this respect, models
derived from microscopic principles are often limited because of the
numerous degrees of freedom strongly coupled to the
magnetization. Hence, phenomenological approaches are often used, and
require a constant comparison with experiments.

\begin{figure}
  \includegraphics[width=7cm]{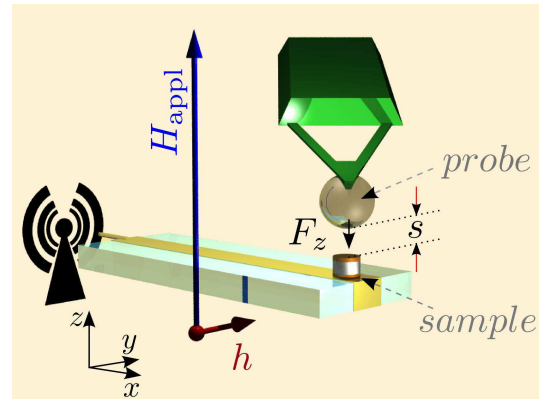}
  \caption{(Color online) Schematic representation of the
    mechanical-FMR setup: the resonance spectra of a small size sample
    are detected by a magnetic force microscope. The probe consists
    here of a magnetic sphere glued at the extremity of the
    cantilever. The excitation comes from a microwave antenna placed
    underneath the sample.}
  \label{mechanical-FMR}
\end{figure}

Several original techniques are being pursued to measure the dynamics
of $\bm{M}$ on small length scales. Among them, X-ray magnetic
circular dichroism (XMCD) transmission microscopy \cite{Stoll04,
  Acremann2006}, XMCD photoelectron microscopy (PEEM) \cite{Vogel03},
microfocus Brillouin light scattering (BLS) \cite{perzlmaier05},
time-resolved scanning Kerr microscopy (TRSKM) \cite{Hiebert97,
  Barman03, Buess04}, spin-torque driven ferromagnetic resonance
(ST-FMR) \cite{tulapurkar05, Sankey2006, Chen08} and magnetic
resonance force microscopy (MRFM) applied to ferromagnetic resonance
\cite{Zhang1996, Charbois02, Loubens07} are the most accomplished yet.
This article focuses on the last one, called herein mechanical-FMR,
since a mechanical setup is used for the detection of FMR, as shown in
Fig.\ref{mechanical-FMR}. Here, we shall review the ability of
mechanical-FMR to extract \emph{quantitative} spectroscopic
information in \emph{individual} samples.

In FMR, the magnitude of the magnetization vector, $\left | \bm M (t,
  \bm r) \right | = M_s$, is a constant of the motion and equals the
saturation magnetization at the lattice temperature. Thus the dynamics
of $\bm{M}$ is governed by the position- and time-dependent transverse
fluctuations of the magnetization direction, the so-called spin-waves
(SW) \cite{Herring51} $\bm m(t, \bm r) = \bm M(t, \bm r)- M_\zeta (\bm
r) \bm \zeta (\bm r)$, where $M_\zeta = \bm M \cdot \bm \zeta$ is the
longitudinal component of the magnetization, defined as the projection
of the instantaneous vector $\bm M(t)$ along the local precession (or
equilibrium) axis of unit vector $\bm\zeta (\bm r)$. Because of the
exchange interaction, these transverse fluctuations are a collective
precession, and become more insightful once decomposed along the
normal modes basis of the sample. Therefore the experimental
identification of this SW basis is important. Its nature depends
primarily on the symmetry of the equilibrium configuration of $\bm M$
inside the sample \cite{Bailleul2006}. In the case of a uniformly
magnetized sample, the SW confinement is mostly governed by the shape
of the sample. In thin films, the quantization of the lowest energy
(or longest wavelength) modes develops along the finite thickness
direction, whereas in nanostructures, it arises both from the
thickness and lateral confinements.

Although, there are fundamental interests in studying the
magnetization dynamics in the time domain (or free precession regime),
the best spectral resolution is obtained by continuous wave
spectrometers, which study the harmonic response (or forced
oscillation regime). Conventional spectrometers \emph{excite} and
\emph{detect} the SW eigen-modes of a ferromagnetic sample with the
same microwave antenna. In a conventional-FMR experiment
\cite{wigen:84}, the sample is placed in a region where a large
homogeneous static magnetic field aligns the spins in a well-defined
direction. On the emission side, a small microwave field $h$, applied
perpendicularly to the effective magnetic field $\bm H_{\rm eff}$
($\parallel \bm \zeta$), will cant the magnetization away from its
equilibrium axis if the resonance condition is met, \emph{i.e.}  if
the photon energy ($\hbar \omega_s$) corresponds to the energy to
excite one SW or magnon ($\hbar \gamma H_{\rm eff}$). The symmetry of
the excitation antenna sets the selection rules and the symmetry of
the excited SW.  On the reception side, the measured quantity in a
conventional spectrometer is $P_\text{abs}$, the microwave energy
absorbed inside the sample per unit of time. Technically,
$P_\text{abs}$ is obtained by monitoring the transmitted or reflected
power to a microwave diode, whose dc voltage is proportional to the
incident microwave power. In metallic samples, the absorbed power per
unit of surface can be expressed as a function of $Z_{s}$, the complex
surface impedance of the sample \cite{oklein:14,oklein:13}, which
itself depends on the complex electrical conductivity and magnetic
susceptibility.  For thin magnetic layers, it mainly reduces to the
dissipative part $\chi''$ of the microwave transverse magnetic
susceptibility, as in insulators: $P_\text{abs}=\int_{V_s} \partial_t
{\bm M} .\bm{h}~dV =\omega_s \chi''h^2 $, where the shorthand notation
$\partial_t {\bm M}$ represents the time derivative.

The dependence of $P_\text{abs}$ on either the applied magnetic field,
$H_{\rm appl}$, or the frequency of the excitation source, $\omega_s/2
\pi$, reveals resonance peaks (bell-shaped curves). Their positions,
amplitudes and line widths reflect the complete spectral information
about the spin system. i) The position is a precise measurement of the
effective magnetic field defined as the conjugate variable of the
magnetization, ${\bm H}_{\rm eff} = \partial_{\bm{M}} {\cal F}$, where
${\cal F}$ is the free energy of the spin system.  ii) The amplitude
is related to the coupling (overlap integral) with both the excitation
and the detection schemes. It thus gives a hint about the
spatio-temporal profile of the mode. For instance, with a cavity
(cavity-FMR) or a stripline antenna (stripline-FMR), the microwave
field $h$ is uniform over the sample volume. As a result, it
preferentially couples to the longest wavelength SW modes of the
sample. iii) The \emph{frequency} line width gives the decay rate of
the excited SW (coherent with $h$) to the other degrees of freedom:
the degenerate and thermal SW (incoherent with $h$) or the lattice
(electrons and phonons). In Bloembergen and Wang's notation
\cite{bloembergen:54}, the line width is proportional to $1/T_2$
\cite{fletcher:955}. Experimentally, it is often easier to monitor the
dependence of $P_\text{abs}$ when $H_{\rm appl}$ is swept at fixed
$\omega_s$.  In this case, it is crucial to repeat the measurement at
several frequencies and to renormalize the \emph{field} line width by
the effective gyromagnetic ratio $\gamma_{\rm eff} =
\partial \omega_s/ \partial {H_{\rm appl}}$.

FMR spectroscopy is usually restricted to probe tiny deviations of the
magnetization from its equilibrium. First, reaching large angles of
precession requires a lot of microwave energy, especially in metallic
samples and second, the interpretation of the high power regime
requires a specific analysis since the equation of motion is
non-linear \cite{suhl:57}.  The typical angles of precession found in
FMR are less than $\vartheta=1 \degree$. Although FMR is a sensitive
technique, still the sensitivity of most conventional spectrometers is
usually not sufficient to detect magnetization dynamics in individual
sub-micron size samples. The crucial parameter here is the ratio
between the volume of the sample and the volume of the detector: the
so-called filling factor. With the recent development of MRFM, the
sensitivity of magnetic resonance has been tremendously
enhanced. Nanometer scale sizes come now within reach as Dan Rugar and
colleagues at IBM Almaden recently showed with the detection of a
single electron spin \cite{rugar04} and with a record 90 nm resolution
on nuclear paramagnets, where only a few thousands of nuclear spins
contribute to the signal \cite{mamin2007a}. The key to such
performance comes from the size of the magnetic probe, which has a
very strong coupling with the resonant volume, thus ensuring an almost
optimized filling factor.

In this paper, we show how a mechanical-FMR setup can be used to
perform FMR spectroscopy of sub-micron size samples. We present a
thorough study of the SW eigen-modes in individual permalloy (Py)
disks, whose FMR spectra are compared to a 2D approximate analytical
model and to full 3D micromagnetic simulations of the dynamical
susceptibility. The structure of the rest of the paper is the
following. Section \ref{MRFM} presents the principles of the
mechanical-FMR and gives an extensive analysis of the spectral
deformations induced by the magnetic probe on the measured FMR
spectra. In section \ref{data}, the experimental results are
presented. In section \ref{models}, we derive an approximate 2D
analytical model and present the results of a 3D micromagnetic
simulation. They are both used in section \ref{comparison} to analyze
the data. Section \ref{conclusion} contains a summary of the results
obtained.

\section{Mechanical-FMR \label{MRFM}}

\subsection{The experimental setup}

\begin{figure}
  \includegraphics[width=7cm]{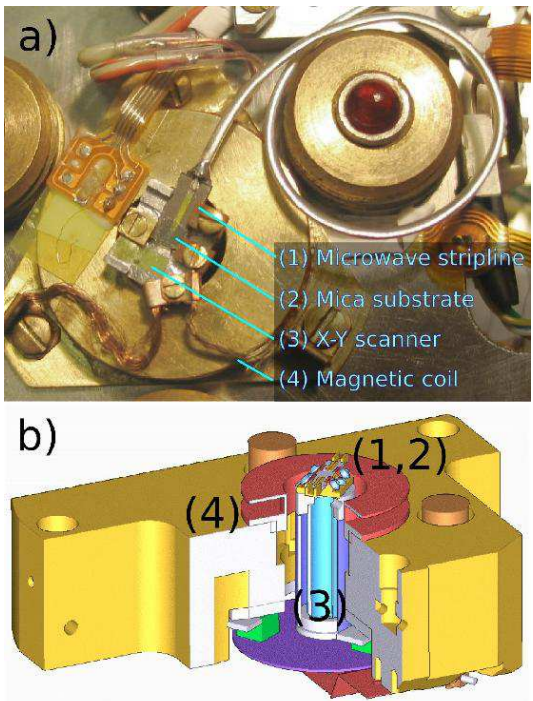}
  \caption{(Color online) a) Photography of the MRFM bottom base
    showing (1) the microwave antenna, (2) the mica substrate where
    the Py disks are, (3) the piezo-electric scanner, and (4) the
    magnetic coil used for a calibrated modulation of the applied
    field. b) Section view of the above elements. The top part with
    the cantilever and the optical detection is not shown.}
  \label{photography}
\end{figure}

The concept of mechanical detection of the magnetic resonance was
first applied to FMR by Phil Wigen and Chris Hammel in 1996
\cite{Zhang1996}. Fig.\ref{mechanical-FMR} illustrates the
mechanical-FMR setup, while Fig.\ref{photography} presents its
experimental realization for the present study.

\subsubsection{Excitation part}

The sample magnetization is excited by a microstrip antenna placed
underneath (see Figs.\ref{mechanical-FMR} and \ref{photography}). The
antenna consists of a $1 ~\mu$m thick gold micro-strip patterned by
optical lithography on top of a sapphire substrate whose bottom
surface is a ground plane (\emph{i.e.} fully covered by Au). In order
to obtain a wide-band antenna, the top Au electrode is shorted to the
ground plane at the extremity of the substrate, hereby creating an
anti-node of microwave magnetic field at the extremity. Placing the
sub-micron Py disks at this location ensures a wide-band excitation
scheme, at least until the magnetic field node, a quarter of the
wavelength away, moves underneath the samples, which occurs at very
high frequency (above 20 GHz). The efficiency of this wide-band setup
to induce large variation of the absorbed power, depends i) on the
amplitude $h$ of the excitation, ii) on the volume of the sample, and
iii) on the susceptibility $\chi"$ of the sample at resonance,
inversely proportional to the line width. It is important to keep the
amplitude of $h$ below saturation or other critical thresholds
\cite{suhl:57}. Indeed, significant spectral distortions are induced
due to non-linear effects. These distortions being themselves a
subject of research \cite{Loubens2005}, all the data shown hereafter
are taken in the linear regime, where the peak amplitude remains
proportional to the excitation power \emph{i.e.}  precession angles
limited to $1 \degree$.

\subsubsection{Detection part}

The detection scheme is directly inspired from magnetic force
microscopy (MFM).

\begin{figure}
  \includegraphics[width=8.5cm]{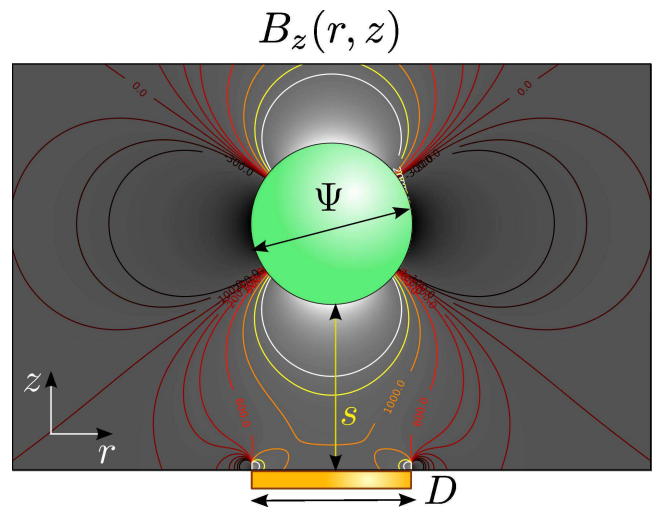}
  \caption{(Color online) Visualization of the dipolar coupling
    interaction between a spherical magnetic probe (diameter $\Psi$)
    and a cylindrical Py disk (diameter $D$) separated by a distance
    $s$. The magnetic configurations in both objects are assumed to be
    saturated along $z$. The color lines represent the iso-$B_z$ field
    lines. The bright and dark shades separate respectively the
    positive and negative coupling regions.}
  \label{iso-Bz}
\end{figure}

A soft cantilever with a magnetic tip is placed in the stray field of
the sample to be studied. The dipolar coupling between the tip and the
sample (see Fig.\ref{iso-Bz}) creates a small flexion of the
cantilever beam, which is detected optically by deflection of a laser
beam on a 4-quadrants photodiode. The pitch angle is produced by the
vertical component of the force $F_z$ (see Fig.\ref{mechanical-FMR})
but also by the torque $N_y$, the component perpendicular to the beam
axis and to the vertical direction. In all the experiments below, we
will measure with a spherical probe the magnetization dynamics of
disk-shaped Py samples perpendicularly magnetized by an homogeneous
applied field. ${H}_{\rm appl}$ is larger than the saturation field of
Py, such that the magnetic state of the sample is almost uniform.  In
this situation, the coupling geometry preserves the axial symmetry,
hence the mechanical coupling comes from the force $F_z$ only
($N_y=0$).

There is a principal difference between the detected signal obtained
by mechanical-FMR and by conventional-FMR. Mechanical-FMR uses
detectors which are only sensitive to $M_\zeta$, the static (or
longitudinal) component of the magnetization, \emph{i.e.} to the
susceptibility averaged on time scales which are much longer than any
relaxation time in the spin system. The microwave oscillations of the
transverse component do not couple to any mechanical mode of the
cantilever, which typically oscillates in the audio range. When an FMR
resonance is excited by the antenna, it is the deviation of $M_\zeta$
from $M_s$, $\Delta M_\zeta = M_s -M_\zeta$, that diminishes the force
$F_z$ on the cantilever (see Fig.\ref{mechanical-FMR}). In a
corpuscular picture, each magnon reduces $M_\zeta$ by $\gamma \hbar$.
Therefore the variation $\Delta M_\zeta$ is proportional to the total
number of magnons being excited by the microwave field inside the
sample, independently of their coherence.  Another way to express this
is through energy arguments. While conventional FMR measures the
\emph{absorbed} power (see above), mechanical-FMR measures the
\emph{stored} energy \cite{Klein2003}.  The ratio between these two
quantities is actually $1/T_1$, the relaxation rate of the
out-of-equilibrium full magnons' population towards the lattice. In
other words, mechanical-FMR provides an intrinsic information about
the dynamics of the SW system, where the area under each resonance
peak of the $\Delta M_\zeta$ spectrum is proportional to $T_1$. But
the $\Delta M_\zeta$ measurement is seldom obtained, because this
quantity is a second order effect in the precession angle
$\vartheta$. $\Delta M_\zeta/M_s = 1 - \cos \vartheta$ is thus much
smaller than the transverse susceptibility, which is of first order in
$\vartheta$.

\subsubsection{Imaging}

Quite naturally, the scanning probe can be used here for imaging
purpose. But the situation is different in paramagnetic \cite{rugar04}
and ferromagnetic spin systems.  In a paramagnet, the spins are
decoupled and the excitation is localized in a ``resonant slice''
\cite{Sidles1995}, whose thickness is inversely proportional to the
field gradient produced by the magnetic tip. In contrast, neighboring
spins in a ferromagnet are coupled through the exchange interaction,
which works against localization effects. The stray field of the
magnetic tip can, however, alter the sample magnetic configuration
just underneath the tip, leading to tip-induced FMR resonances. The
group of Chris Hammel has recently showed, that it is possible to
produce these new localized FMR modes by approaching the magnetic tip
close to the sample surface \cite{obukhov08}. Exploitation of the
spectral features of these new modes is still a challenge at the
moment.

Detection wise, one benefits from the same advantage of an MFM. The
spatial resolution is related to the size of the magnetic probe and
the separation with the sample. In the present study, we are mostly
interested by an optimization of the sensitivity of the mechanical
detection, allowing us to measure smaller sample sizes. As will be
shown below, optimization of the sensitivity requires to chose a size
of the magnetic probe of the order of the size of the sample (optimum
filling factor \cite{oklein:04}). Furthermore, placing the probe far
away from the sample surface and working in the weak coupling regime
diminishes the spectral alteration produced by the tip. Such
conditions are quite obviously incompatible with good imaging
conditions and an increase of the spatial resolution (smaller probe)
must then come at the detriment of the sensitivity.

\subsection{Detection sensitivity}

One of the main advantage of the mechanical-FMR is its exquisite
sensitivity. This is mainly due to the progress in nanofabrication
technologies, which can produce micron-size mechanical structures with
outstanding performance figures, \emph{i.e.} achieving among the best
compromise between small size and large quality factor.

\subsubsection{Modulation technique}

Exciting the sample at a fixed frequency ($\omega_s/2\pi$),
spectroscopy is achieved by recording the cantilever motion as a
function of the perpendicular dc applied field, $H_{\rm appl}$,
produced by an electromagnet. As mentioned before, the coupling
between the mechanical oscillator and the microwave magnetization
dynamics is purely static. However, it is possible and useful to
modulate the microwave power at the mechanical resonance frequency of
the first flexural mode of the cantilever $f_c= 3$ kHz in our
case). As a result, the amplitude of vibration will be multiplied by
$Q$, the quality factor of the mechanical resonator. Note that the
mechanical noise is also amplified. However, a sensitivity gain is
obtained if this intrinsic mechanical noise exceeds the preamplifier
noise or the noise of the microwave source. This modulation technique
is referred to source or amplitude modulation.  The amplitude of the
microwave field follows the time dependence
\begin{equation}
  h(t)=h e^{i\omega_st} \left\{ \frac{1}{2}+\frac{1}{2}\cos(2 \pi f_c t) \right\},
  \label{sourcemod}
\end{equation}
where the depth of the modulation is 100\%. Note that this modulation
technique does not affect the line shape in the linear regime, because
the period of modulation $1/f_c$, is very large compared to the
relaxation times $T_1$ and $T_2$ of the ferromagnetic system studied.

We mention that the modulation of the microwave field at $f_c$ also
induces a direct vibration of the cantilever, even outside any
resonance phenomena of the probe or of the sample. We attribute this
to a modulation of the temperature of the cantilever. The latter is a
direct consequence of the modulation of the microwave heating and eddy
currents, mainly induced by $e$, the electric component of the
electromagnetic radiation. This effect can distort the resonance peaks
due to the modulation of the applied field on the sample by the
vibrating magnetic probe. For instance, a sphere with a magnetic
moment of $6~10^{-9}$ emu vibrating by 10 nm$_{\text{pp}}$ induces a
field modulation of about 6 Oe$_{\text{pp}}$ on a sample placed at
$2.7~\mu$m from its center (separation $s=1.0$ $\mu$m, see
below). This field modulation has no influence on the FMR signal if it
is small compared to the line width, but will otherwise significantly
broaden the resonance. To eliminate it, a forced oscillation out of
phase with the spurious contribution is produced by a piezo-electric
bimorph slab placed nearby the cantilever. The resulting total
vibration corresponds to a few nanometers at most. All the spectra
shown below are obtained in these conditions, where the amplitude of
vibration of the cantilever is compensated so that it has no influence
on the measured FMR spectra.

\subsubsection{Minimal detectable force}

In all MRFM setups, the detection noise is only limited by the
Brownian motion of the cantilever, which behaves as an harmonic
oscillator with a single degree of freedom. The theorem of
equipartition of energy stipulates that, in the absence of external
force, the vibration amplitude of the cantilever is such that the
average kinetic energy associated to the thermal excitation is equal
to $k_B T/2$, where $k_B$ is the Boltzmann constant and $T$ is the
temperature of the bath. Thus the minimal detectable force follows the
relation
\begin{equation}
  F_{min}=\sqrt{\frac{2 k_B T k B}{\pi f_c Q}},
  \label{fmin}
\end{equation}
where $k$ is the spring constant of the cantilever, $Q$ its quality
factor, and $B$ the detection bandwidth. FemtoNewton sensitive
cantilevers are now readily available commercially. A BioLever B from
Olympus with $k=5~10^{-3}$ N/m has been used for this work. Because
the cantilever is very sensitive to thermal fluctuations, it is
important to stabilize both the intensity of the laser (to less than
15 ppm) that is shined on top of the cantilever for the position
sensing but also the over-all temperature of the microscope (to less
than 200 ppm), that is mounted on Peltier elements. To increase the
sensitivity, the experiment is operated in a secondary vacuum of about
$10^{-6}$ torr. Operating in vacuum is important to keep the large
value of $Q=4500$. As seen in Eq.(\ref{fmin}), the other parameters
that control the minimum detectable force are the temperature and the
bandwidth. Working at low temperature and averaging the signal over
large period of time allows the detection of attoNewton
forces\cite{Mamin2001}. All the data shown hereafter are obtained at
$T=280$ K and with a lock-in time constant of one second.

\subsubsection{Magnetic spherical probe}

\begin{figure}
  \includegraphics[width=8.5cm]{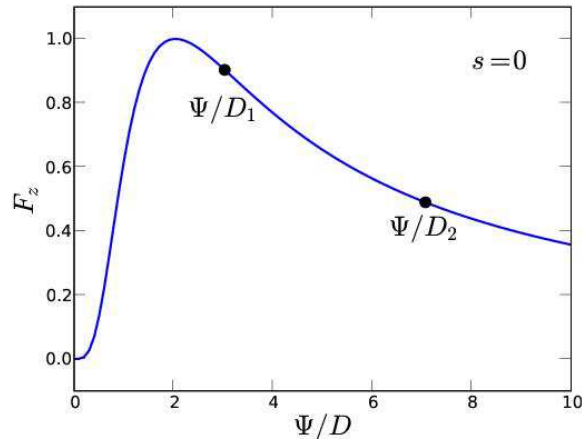}
  \caption{(Color online) Variation of the dipolar magnetic force
    between a sphere and a disk in contact ($s=0$) as a function of
    the ratio between their diameters, respectively $\Psi$ and $D$.
    The maximum force occurs when $\Psi=2 D$.}
  \label{optimalPsi}
\end{figure}

Obviously the sensitivity directly depends on the strength of the
dipolar coupling between the probe and the sample. As mentioned above,
the magnetic probe must be carefully chosen, because there is an
optimal size for a given sample (concept of filling factor). In our
case, the probe itself is a magnetic sphere glued at the apex of the
cantilever. For a spherical probe, the static force applied on the
cantilever is simply proportional to $g_{zz}$, the $z$ component of
the field gradient created by the sample at the center of the sphere:
\begin{equation}
  F_z={m}_{\rm sph}g_{zz},
  \label{forcesph}
\end{equation}
where ${m}_{\rm sph}$ is the magnetic moment of the sphere. For a
disk-shaped sample, the largest force is obtained when the radius of
the sphere is equal to the diameter of the disk \cite{klein04}, as
shown in Fig.\ref{optimalPsi}. The optimum corresponds to the
particular case where the sphere captures all the positive field lines
emanating from the disk (see Fig.\ref{iso-Bz}).

\begin{figure}
  \includegraphics[width=7cm]{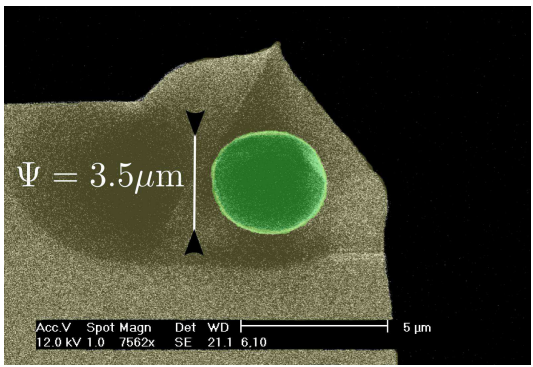}
  \caption{(Color online) Scanning Electron Microscopy image of the
    CoFeSi magnetic sphere glued at the tip of an Olympus Biolever.}
  \label{sphere}
\end{figure}

A Scanning Electron Microscopy (SEM) image of the sphere is shown in
Fig.\ref{sphere}. Its diameter is $\Psi=3.5~\mu$m, which is slightly
larger than the optimal size for our $D_1= 1.0$ and $D_2=0.5~\mu$m
sample diameters. This sphere is an amorphous alloy whose main
constituents are Co (80 wt\%), Fe (10 wt\%) and Si (9 wt\%), as
deduced from chemical analysis. Its characteristics are measured after
the gluing process of the sphere at the tip of the cantilever. The
magnetization curve is obtained by placing the mounted cantilever
above an Fe cylinder (diameter 2 mm, height 8 mm), which creates a
well-characterized field gradient of $g_{zz}=0.5$ G/$\mu$m. By
monitoring the deformation of the cantilever vs. the applied field, we
can infer a 1.2 kG saturation field for the sphere. We also obtain the
value of its magnetic moment ${m}_{\rm sph}=(6\pm0.5) 10^{-9}$ emu,
which is deduced from the known values of the field gradient and the
cantilever's spring constant.

We have already mentioned above the interest to preserve the axial
symmetry, which is achieved if the center of the sphere is positioned
above the center of the disk. Two other comments should be added about
the choice of the spherical shape for the magnetic probe. First, in
contrast to cylinders, spheres do not have any shape anisotropy. This
property, combined with the absence of magneto-crystalline anisotropy
of the amorphous alloy, considerably decreases the influence of the
applied magnetic field on the mechanical resonance frequency of the
cantilever. This is important because the source modulation technique
relies on a phase locked loop (PLL) to keep the source modulation
exactly at the resonance frequency of the cantilever. Second, the
magnetic probe on the cantilever is itself placed in the field
gradient of the sample. As a result, mechanical vibrations of the
cantilever are also produced by the excitation of the FMR modes of the
probe itself. This can be seen as a reciprocal effect of the
probe-sample coupling. Using different shapes for the probe (sphere)
and for the sample (disk), the two FMR spectra are completely
separated, mainly due to the difference in the shape anisotropies of
the two different geometries. In the spectral range shown in this
study ($\omega_s/2 \pi=4$ to 10 GHz), the FMR resonance of the sphere
occurs below 3 kOe \cite{Loubens2005a}, \emph{i.e.} at much lower
field than for the disks studied hereafter.

\subsection{Extrinsic effects}

\subsubsection{Spectral deformations induced by the magnetic probe}

The optimization of the sensitivity discussed above has, however, a
drawback. Increasing the coupling between the probe and the sample
inevitably produces a distortion of the spectral features. It is thus
important to understand this effect and to keep it inside the
perturbation regime so that one could still deduce the intrinsic
behavior of the sample, \emph{i.e.} without the presence of the probe.
In mechanical-FMR, the spectral distortions are due to the stray field
of the probe. It produces an additional field inhomogeneity in the
internal effective magnetic field, which affects the detailed SW
dispersion relation inside the sample and thus the resonance
condition. The strength of the effect depends on ${m}_{\rm sph}$ and
on the separation $s$ between the probe and the sample (see
Figs.\ref{mechanical-FMR} and \ref{iso-Bz}). We discuss in the
following how to ensure that the field inhomogeneity produced by the
probe is small compared to the internal dipolar field variations along
the radial direction of the sample ($\approx 2 \pi M_s$ for a
non-ellipsoidal shape).

Thorough experimental studies have been performed in the past on
mechanical-FMR of a Y\(_{3}\)Fe\(_{5}\)O\(_{12}\) single crystal disk
for different separations $s$ between the magnetic tip and sample
surface \cite{Charbois02}. It was found that, when the bias field
inhomogeneity from the probe is smaller than a few percents of the
internal field variation inside the sample, its main effect is to
shift the entire spectrum to higher frequency as $s$ decreases.  This
shift is homogeneous within less than 10\% for all the peaks
corresponding to the different SW modes. In these conditions, their
spatial profiles are almost not affected by the presence of the probe,
and the relative amplitudes between the peaks are kept to their
intrinsic values. On the other hand, if the bias field from the tip
represents a large perturbation to the internal field inside the
isolated sample, the spatial profiles of the SW modes are affected
\cite{urban06, Loubens2005a}, and the observed FMR spectrum can not be
recognized as intrinsic to the sample. This situation can also lead to
a different coupling between the probe and the sample
\cite{Loubens2005a}. However, it is to note that since the mechanical
probe is independent from the excitation, an increase of the coupling
between the sample and the probe do not produce any additional
contribution to the FMR line width \cite{bloembergen:54}.  This is in
contrast to the radiation damping found when the coupling to the
microwave resonator increases \cite{Klein2003}.

\subsubsection{Quantitative analysis \label{shift}}

A quantitative estimation of $H_{\rm off}$, the shift induced by the
tip on each SW mode, can be obtained analytically within a 2D model.
By 2D, we mean here a model where the precessional profile of each
mode in the disk only depends on the two in-plane coordinates,
\emph{i.e.} on $(r,\phi)$ in a cylindrical frame (it is uniform along
the thickness direction $z$). In that case, the resonance condition
depends on the phase delay of the SW accumulated along the
diameter. Constructive interferences occur when the phase delay over a
cycle is equal to $n \times 2 \pi$, $n\in\mathbb{N}^*$. This condition
is equivalent to the WKB approximation applied to the dispersion
relation of magnetostatic forward volume waves (MSFVW), established by
Damon and Eshbach in 1961 \cite{Damon1961}. In the following, we will
use the dipole-exchange dispersion relation developed by Kalinikos and
Slavin \cite{Kalinikos1986} for magnetized thin films and later
applied by Kakazei \emph{et al.} for the case of disks magnetized in
the exact perpendicular geometry \cite{Kakazei2004}.

\begin{figure}
  \includegraphics[width=8.5cm]{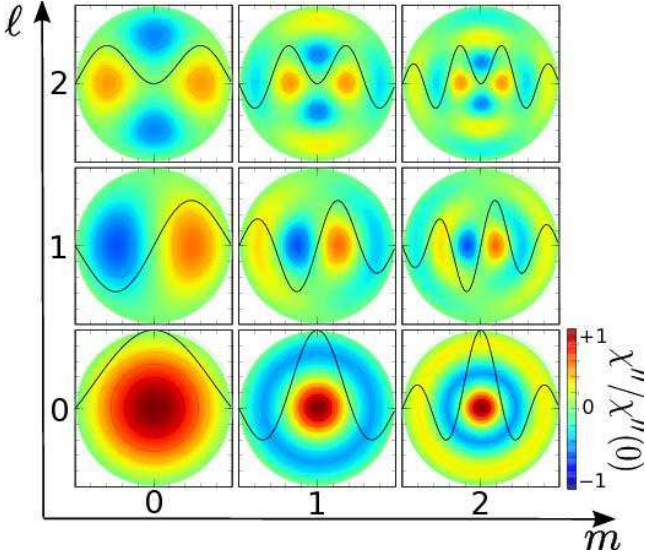}
  \caption{(Color online) Color code representation of the eigen-modes
    basis ${\cal J}^{\ell}_m (\bm r)$, where $\ell$ and $m$ indicate
    respectively the number of nodes in the circumferential and radial
    direction. Only the modes $\ell=0$ couple to a uniform microwave
    excitation.}
  \label{eigen-modes}
\end{figure}

We start with the normal modes basis of a disk magnetized in the
perpendicular direction, which are the ${\cal J}^{\ell}_m (\bm r) =
J_\ell (k_{\ell,m} r)\cos(\ell \phi)$, where $J_\ell$s are the Bessel
functions of the first kind and $k_{\ell,m}$ is the modulus of the
in-plane SW wave vector determined by the boundary conditions. We
shall assume that the modes satisfy the dipolar pinning condition at
the circumference of the disk samples, \emph{i.e.} ${\cal J}^{\ell}_m
(\bm R)=0$. Thus $k_{\ell,m} = \kappa_{\ell,m}/R$, where
$\kappa_{\ell,m}$ is the $(m+1)^\text{th}$ root of $J_\ell(x)$ and $R$
is the radius of the disk. In this notation, $\ell$ and $m$ are
respectively the azimuthal and radial mode indices (\emph{i.e.} the
number of nodes in the circumferential and radial
directions). Fig.\ref{eigen-modes} is a color-coded representation of
the transverse susceptibility $\chi"$ corresponding to the first
modes.

The obtention of an analytical formula for the resonance condition
shall be detailed in section \ref{analmod} for the general case where
the applied magnetic field makes a small angle $\theta_H$ with the
normal of the disk (see Eqs.(\ref{tyb-arb-theta-1}) and
(\ref{tyb-arb-theta-2})). For the estimation of the spectral
deformations induced by the probe, we are only interested by the
$\theta_H=0\degree$ limit of the linearization of the equation of
motion. The resonance condition for the normal mode $(\ell,m)$ in the
perpendicular direction is given by the expression
\cite{Kalinikos1986}:
\begin{eqnarray}
  \omega_s & = & \sqrt{ \left( \omega_{\rm int} + \omega_M \Lambda^2
      k_{\ell,m}^2 \right) } \times \nonumber \\
    & & \sqrt{ \left( \omega_{\rm int} + \omega_M \Lambda^2  k_{\ell,m}^2 + (\frac 12 - 
      G_{\ell, m}^{\perp} )\omega_M  \right) }.
  \label{omega_s}
\end{eqnarray}
Here, $\omega_M = 4 \pi \gamma M_s$ and $\omega_{\rm int} = \gamma
H_{\rm int}$ is the mode dependent effective internal magnetic field
\begin{equation}
  H_{\rm int}=H_{\rm ext} - \widetilde{N_{zz}} 4\pi M_s + H_{\rm A}, 
  \label{Hi}
\end{equation}
where $H_{\rm ext}$ is the total external magnetic field,
$\widetilde{N_{zz}}$ is the longitudinal matrix element of the
effective demagnetizing tensor (see Eq.(\ref{N-dyn})) and $H_{\rm A}$
is the perpendicular uniaxial anisotropy field (of spin-orbit coupling
origin). $\Lambda=\sqrt{2 A/(4\pi M_s^2)}$ is the exchange length,
which depends on the exchange stiffness constant $A$, expressed in
erg/cm ($=10^{-6}$ in Py). Finally, the parameter $G_{\ell,
  m}^{\perp}$ is derived in section \ref{analmod}, Eq.(\ref{G-perp}).

In our notation, $\hat{\bm N}$ is the demagnetizing tensor of the
disk. Appendix \ref{appA} gives the analytical expression for the
different matrix elements of this tensor in the cylindrical
coordinates. In Eq.(\ref{Hi}), the demagnetizing field along $z$
depends on a matrix element of the effective demagnetization tensor
$\hat{\widetilde{\bm N}}_{\ell,m}$. The latter is the demagnetizing
tensor weighted by the spatial dependence of the normal mode profile
${\cal J}^{\ell}_m ( \bm r)$,
\begin{equation}
  \hat{\widetilde{\bm N}}_{\ell,m} = \frac{1}{C_{\ell,
      m}}\int_{r < R} \hat{\bm N}(\bm r) {\cal J}^{\ell}_m (\bm r)^2 d^2\bm r,
  \label{N-dyn}
\end{equation}
where $C_{\ell,m}$ is a renormalization constant:
\begin{equation}
  C_{\ell,m} = \int_{r < R} {\cal J}^{\ell}_m (\bm r)^2 d^2\bm r.
  \label{Clm}
\end{equation}

To calculate the influence of the sphere on the resonance field of the
mode $(\ell,m)$, we need to expand the external field as the sum of
two contributions: the homogeneous magnetic field $H_{\rm appl}\bm z$
produced by the electromagnet and the inhomogeneous stray field of the
sphere $\bm H_{\rm sph}(\bm r)$:
\begin{equation}
 \bm H_{\rm ext}(\bm r) = \bm H_{\rm appl} + \bm H_{\rm sph}(\bm r).
\end{equation}
Here we are only interested by its $z$-component, $H_{\rm sph}(\bm r)
= \bm H_{\rm sph}(\bm r)\cdot \bm z$. Then, the influence of the probe
simply yields a modification of the internal magnetic field $H_{\rm
  int}$ of Eq.(\ref{Hi}) by
\begin{equation}
  H_{\rm int} = H_{\rm appl} + H_{\rm off} - \widetilde{N_{zz}}
  4\pi M_s + H_{\rm A},
  \label{hi-complet}
\end{equation}
where the shift $H_{\rm off}$ induced by the probe is averaged along
the radial direction by the mode profile:
\begin{equation}
  H_{\rm off}=\frac{1}{C_{\ell,m}}\int_{r< R}H_{\rm sph}(\bm r) {\cal J}^{\ell}_m(\bm r)^2d^2\bm r.
  \label{hoff}
\end{equation}
This expression allows us to give an estimation of the maximum
coupling allowed to keep the spectral deformations in the perturbation
regime. The additional contribution $H_{\rm off}$ must be small
compared to the internal field variation inside the sample volume
($\sim\widetilde{N_{zz}} 4\pi M_s$) in order to keep the normal mode
basis unchanged.  In the opposite case, the SW mode profiles and their
resonance fields must be directly calculated from the MSFVW dispersion
relation and from the magnetostatic potential. Clear experimental
signatures of this strong influence of the probe on the SW mode can be
found \cite{Loubens2005a}. The experiments presented below are not in
this regime.

\begin{figure}
  \includegraphics[width=8.5cm]{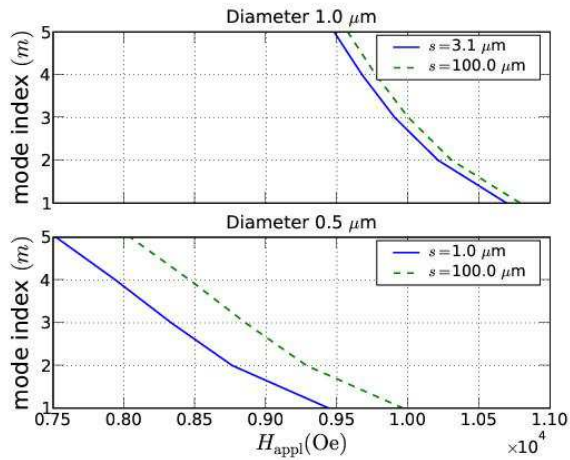}
  \caption{(Color online) Calculated spectral deformation induced by
    the magnetic tip placed at distances $s = 3.1$ and $1.0~\mu$m
    above the disk samples $D_1=1.0$ and $D_2=0.5~\mu$m,
    respectively. The graphs show the downward shift in field of each
    resonance mode (indexed by $m$) compared to the unperturbed FMR
    spectrum (dashed line).}
  \label{probe-pertubation}
\end{figure}

Quantitative results of the calculation are displayed in
Fig.\ref{probe-pertubation} for the two different disks diameters,
$D_1=1.0$ and $D_2=0.5~\mu$m. We have calculated the influence of the
probe on the circumferentially symmetric ($\ell=0$) normal mode ${\cal
  J}^{0}_m (\bm r)$. The value of $M_s$ and $\gamma$ used in the
calculation are those derived below from cavity-FMR studies of the
extended thin film. We also use the physical characteristics of the
probe derived from the SEM images in Fig.\ref{sphere} and
magnetometry. The values of $s$ (the separation between the sphere and
the disk) used in Fig.\ref{probe-pertubation} are close to the
experimental conditions.  The magnetic probe is brought closer to the
disk surface as the sample diameter diminishes in an attempt to
compensate for the reduction in the coupling strength. The induced
force is expected to decrease by approximately an order of magnitude
between the two disks diameters: a factor of 4 comes from the decrease
of the resonating volume and a factor of 2 comes from the decrease of
the filling factor (see dots on Fig.\ref{optimalPsi}). On the other
hand, approaching the probe from $s=3.1$ to $1.0~\mu$m corresponds to
an increase of the gradient by an order of magnitude (from $g_{zz} =
6.1 \times 10^5$ to $5.8 \times 10^6 $~G/cm, respectively). For
comparison, the dashed line is the $s=100~\mu$m case, which represents
the intrinsic, unperturbed case.  Two important observations can be
deduced from these plots.  First, the shift is almost independent of
the mode number $m$. The spectral deformation can thus simply be
modeled by an offset on the applied field. Second, at constant
coupling, the amplitude of this offset increases when the disk
diameter decreases, as the probe has to be brought closer to the
smallest sample. In the case of our magnetic sphere, the displacement
of the resonance field is about $-100$ Oe downward in field for the
large disk and of $-520$ Oe for the small one.

In conclusion, under experimental conditions of an almost optimized
coupling between the sample and the probe, the influence of the latter
on the intrinsic FMR spectrum of the former is only an over-all
downward shift in field, which can be quantitatively estimated. The
fact that our sphere has a small moment and a large diameter keep the
stray homogeneous over the sample volume and reduces deformation in
the relative position of the resonance modes. Assuming that the sphere
is brought into contact with the surface of the disk, the variation of
the perpendicular stray field along the radial direction of the disk
is about 400 Oe, which is small compared to the $\widetilde{N_{zz}} 4
\pi M_s \approx 10^4$ Oe variation of the internal field inside the
disk.

\section{Spectroscopy of sub-micron size disks \label{data}}

\subsection{Film layered structure}

Films corresponding to the composite system (permalloy $=$
Ni$_{80}$Fe$_{20}$, abbreviated by Py $|$ alumina) have been deposited
\cite{Hurdequint2002} by RF sputtering, at room temperature, on two
different substrates: single crystalline Si and mica. A sweeping mode
for sputtering has been used, where for each material the
substrate-holder is swept back and forth over the activated
target. Such a mode of deposition was chosen to achieve a better
homogeneity of the magnetic layer and a good control of its
thickness. Several multilayers (Py $|$ Al$_2$O$_3$)$_N$ have been
produced, using this sweeping mode, being characterized by an
individual layer in the ultrathin range. Low-angle X-ray diffraction
measurements have been performed on these multilayers (Si substrate)
and the exploitation of these results has led to the determination of
the deposition rates for the two materials. The specific magnetic film
studied in the present report has the following layered structure:
(Al$_2$O$_3$ base $|$ Py $|$ Al$_2$O$_3$ top). It consists of a single
Py layer (43.3 nm thick) sandwiched by two Al$_2$O$_3$ layers of
identical thickness (16 nm). The top alumina layer protects the Py
layer from oxydation. The smooth amorphous Al$_2$O$_3$ base layer,
characterized by a small surface roughness (a few \AA) helps the
growth of (111) textured Py polycrystalline layers in the low
thickness range for the Py layers. The typical spread of surface
orientation of the crystallites in the (111) textured Py layer grown
by sputtering on the amorphous alumina is $\Delta \theta_H=0.5
\degree$. Due to the symmetry of the multilayer structure, the
pinnings of the magnetization at the top and bottom (Py $|$
Al$_2$O$_3$) interfaces are expected to be identical. Most importantly
it is to be noted that, in this layered structure, the Py layer is
sandwiched by an insulator. Consequently, by contrast to the case
\cite{Hurdequint2007,mizukami1} of an adjacent metallic layer (Py $|$
N), where N designates a normal metal, no modification of the Py
intrinsic damping (a spin-diffusion phenomena) is expected to arise
from the (Py $|$ Al$_2$O$_3$) interfaces.

\subsection{Cavity-FMR studies of the extended thin films}

Cavity-FMR experiments have been performed on the extended thin films
deposited on Si and mica, corresponding to the layered structure
described above (16 Al$_2$O$_3$ $|$ 43.3 Py $|$ 16 Al$_2$O$_3$), with
individual thickness in nm. The basic FMR experiment which has been
carried out (reflexion at X-band, 9.6 GHz, and room temperature)
consists in studying the resonance spectrum as a function of the
orientation $\theta_H=({\bm z}, {\bm H_{\rm appl}})$ of the dc field
$H_{\rm appl}$ applied in a plane perpendicular to the film. The
resonance condition \cite{Hurdequint2002} for the uniform mode, as a
function of $\theta_H$, depends only on two parameters: the
gyromagnetic ratio $\gamma$ and the total perpendicular anisotropy
field, $H_u = 4 \pi M_s-H_{\rm A}$. The two fitting parameters
($\gamma, H_u$) are deduced from the observed angular variation of the
resonance field. Very close values of these two parameters are found
for the the films grown on Si and on mica. For the film grown on mica,
used to pattern the sub-micron disks, the gyromagnetic ratio is
$\gamma=1.849~10^{7}$ rad.sec$^{-1}$.Oe$^{-1}$, corresponding to a
Land\'e factor $g=2.103\pm0.004$.  The perpendicular anisotropy field
$H_u = 9775 \pm 50$ Oe reflects entirely the demagnetizing field $4
\pi M_s$ and corresponds to the expected value of the magnetization
for a NiFe alloy of atomic composition Ni$_{80}$Fe$_{20}$. In fact,
the spin-orbit anisotropy field $H_A$, that is ascribed
\cite{Hurdequint2002} to a stress-induced anisotropy observed in the
ultra-thin NiFe layers, is here, for this thick Py layer, nearly zero
(at least $< 100$ Oe) due to the fact that for this alloy composition
the magnetostriction reduces nearly to zero. We mention also that this
thick Py layer is characterized by a small uniaxial in-plane
anisotropy, which is identified to a field-induced anisotropy built
during the film deposition. Its small value, $7.5 \pm 0.5$ Oe, is
deduced from the two parallel geometry FMR measurements with the
magnetic field applied respectively along the easy and the hard axis.

\subsection{Mechanical-FMR studies of the patterned structures}

\begin{figure}
  \includegraphics[width=7cm]{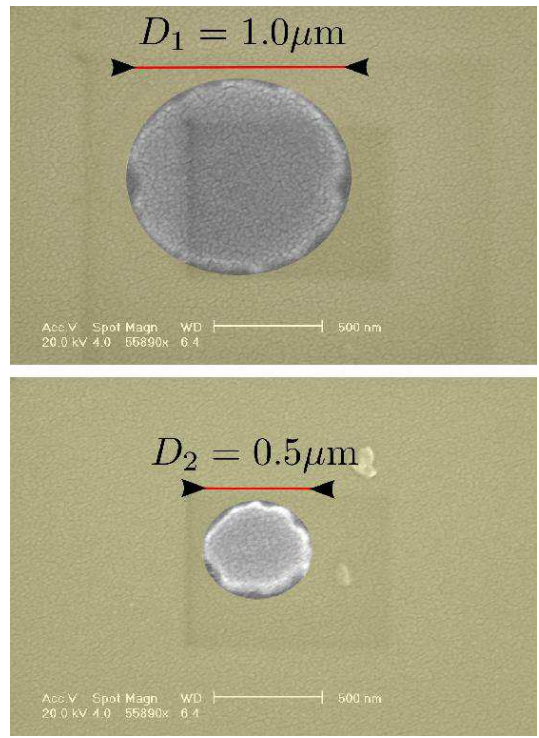}
  \caption{(Color online) SEM images of the two Py disks samples. The
    images are being viewed at an angle, thus distorsions exist in the
    vertical direction. }
  \label{disks}
\end{figure}

The disks prepared for the mechanical-FMR experiment were patterned
out from the (Al$_2$O$_3$ $|$ Py $|$ Al$_2$O$_3$) film grown on
mica. Because this substrate cleaves easily, it was possible to reduce
its thickness down to about $15~\mu$m before gluing it on the
broadband stripline.  As a result from the short distance between the
sample and the excitation circuit, the ac field $h$ at the sample
location can achieve up to 10 Oe for the available power from the
synthesizer.  However, in the results presented below, $h$ was
restricted to 1 Oe in order to avoid non-linear effects. The
mechanical-FMR data are all collected at $T=280$ K. The microscope was
aligned so that the normal of the mica substrate was lying within $5
\degree$ with the dc field.

The disks studied below have been patterned out of the same thin film
deposited on mica which was studied by cavity-FMR. An aluminium mask
defined by e-beam lithography is used to protect the Py disks during
the subsequent ion etching of the thin magnetic film. Several disks of
nominal diameters 0.5 and 1.0 $\mu$m separated from each other by 50
$\mu$m were defined using this lithographic process. Fig.\ref{disks}
shows SEM images of two such patterned disks. They do not correspond
to the measured disks, whose FMR spectra are presented in
Figs.\ref{figdisk1} and \ref{figdisk2}, but were processed using the
same recipe. A thin Au layer had to be deposited before the SEM
imaging to avoid charging too much the mica substrate. The shape and
the dimensions of the samples are as expected (the slightly elliptical
shape seen on the images is due to perspective). The lateral rugosity,
which is about 30 nm, could be due to uncertainties in the
lithographic process on mica, or to an inappropriate dose during the
insulation. It is not expected to influence too much the eigen-modes
profiles of a perfect disk (presented in Fig.\ref{eigen-modes}) and
their resonance fields as their amplitude vanish at the periphery of
the disk.  These radial fluctuations do not either contribute to the
FMR line width when the measurement is made with the field parallel to
the normal of the disk. In that case, only inhomogeneities along the
thickness (\emph{i.e.} the magnetization direction) have a direct
impact on the line width broadening \cite{wigen:84}.

\subsubsection{Mechanical-FMR study of the $D_1=1.0~\mu$m disk}

\begin{figure}
  \includegraphics[width=8.5cm]{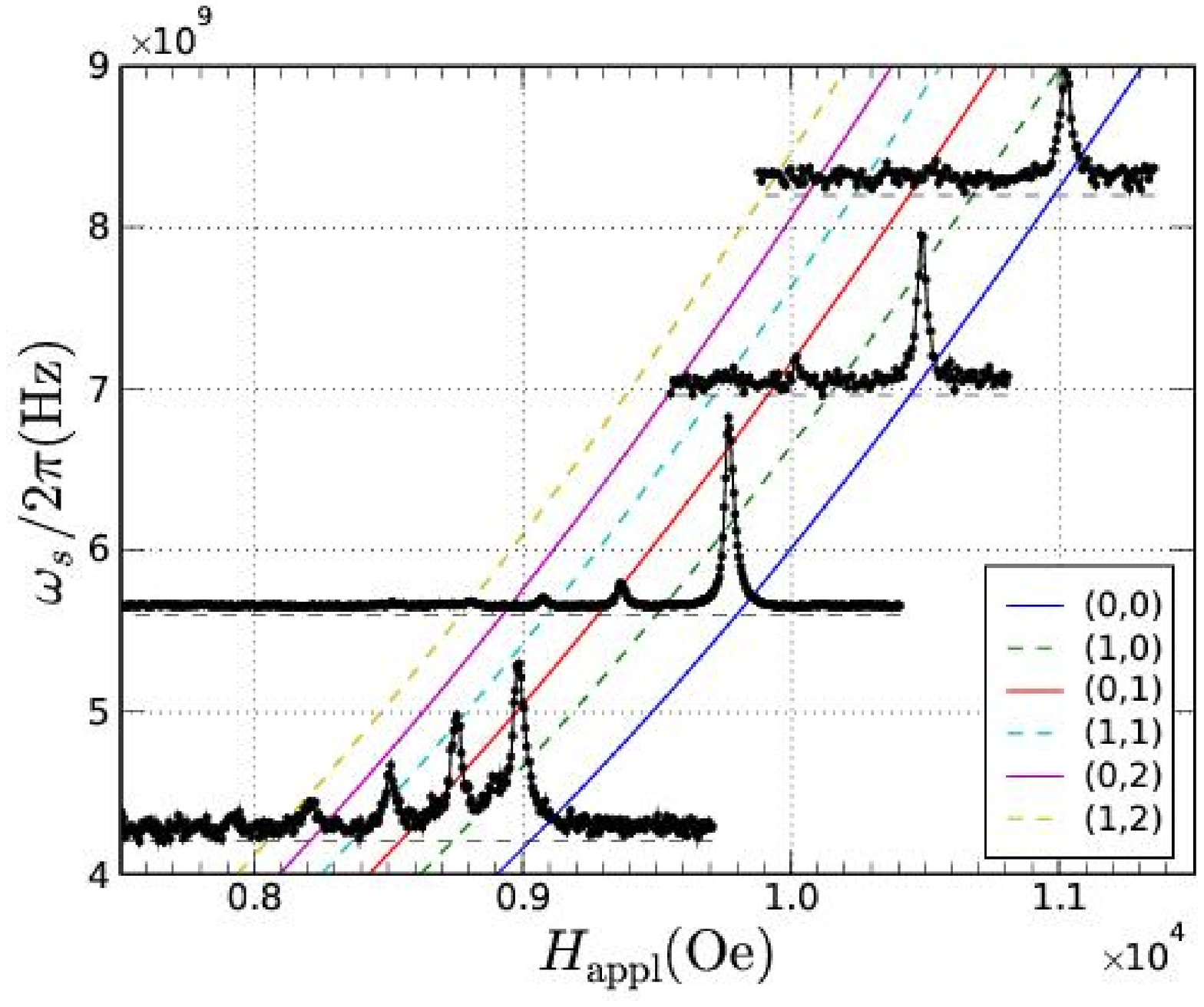}
  \caption{(Color online) Mechanical-FMR spectra of the $D_1=1.0~\mu$m
    disk measured at 4.2, 5.6, 7.0 and 8.2 GHz and $s\approx3.1~\mu$m.
    The lines are the analytical predictions of the locus of the
    $(\ell,m)$-modes when $\theta_H = 4.9 \degree$ and $H_{\rm off} =
    -70$ Oe.}
  \label{figdisk1}
\end{figure}

The mechanical-FMR measurement on the $D_1=1.0~\mu$m disk was
performed with a probe-sample separation $s\approx3.1~\mu$m, kept
constant at the four different frequencies studied and shown in
Fig.\ref{figdisk1}.  To ensure that the separation is the same between
the different spectra, a fine tuning of $s$ is operated at the
beginning of each scan (same $H_{\rm appl}$) so as to keep the
frequency of the cantilever identical.  A series of magnetostatic
modes is observed at each microwave frequency, with the most intense
at the highest field, \emph{i.e.} at the lowest energy.  It
corresponds to the mode whose coupling with the excitation is maximum,
because it is the longest wavelength mode. We note that the 5.6 GHz
spectrum has been measured for a larger integration time than the
other spectra, which explains its better signal to noise ratio and
allows to clearly observe the small amplitude low field modes.  The
detailed analysis of the spectral features (position, amplitude and
line width) of the $D_1=1.0~\mu$m disk is done in section
\ref{comparison}, based on the models developed in section
\ref{models}.

\subsubsection{Mechanical-FMR study of the $D_2=0.5~\mu$m disk}

\begin{figure}
  \includegraphics[width=8.5cm]{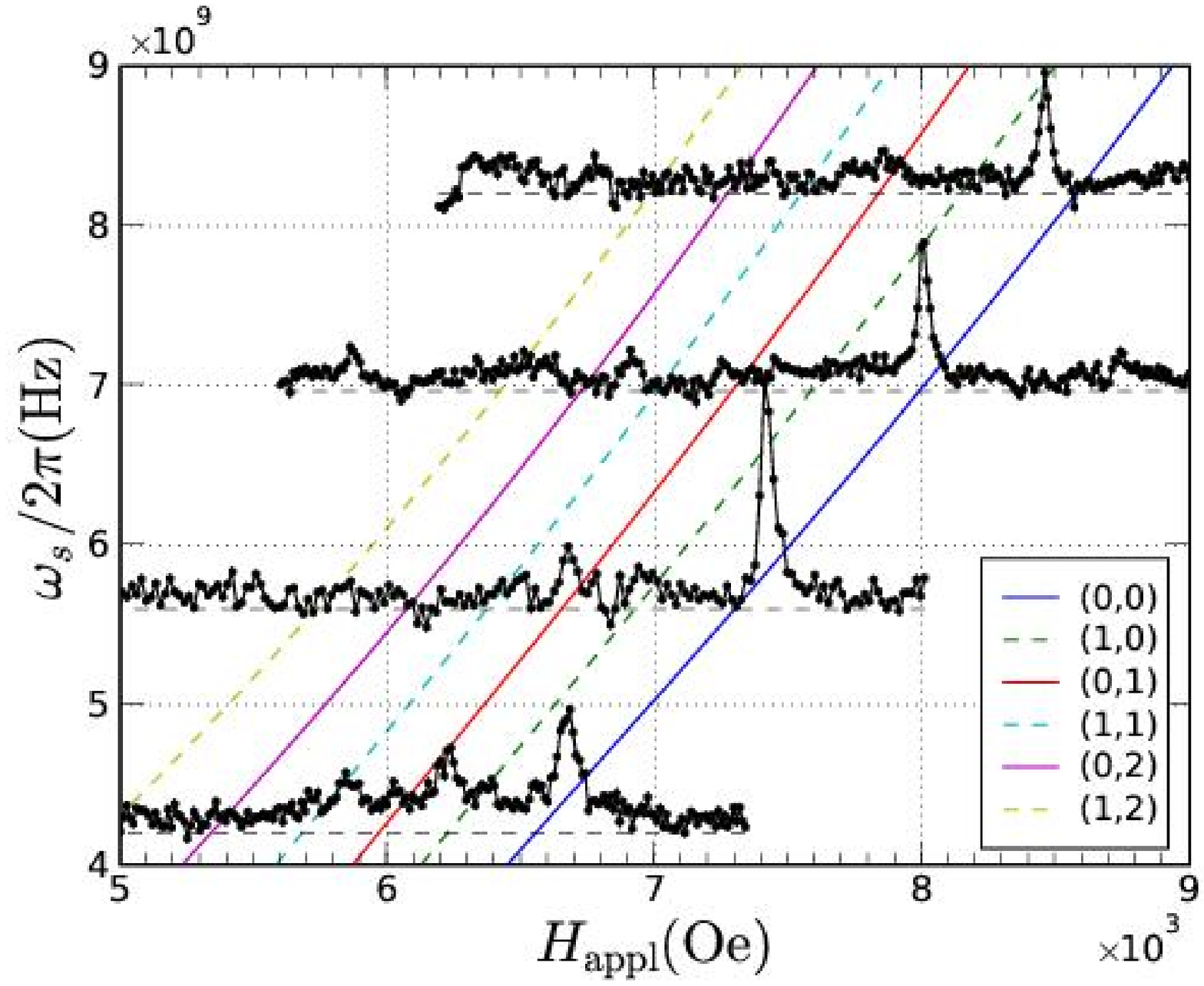}
  \caption{(Color online) Mechanical-FMR spectra of the $D_2=0.5~\mu$m
    disk measured at 4.2, 5.6, 7.0 and 8.2 GHz and $s\approx1.0~\mu$m.
    The lines are the analytical predictions of the locus of the
    $(\ell,m)$-modes when $\theta_H = 9.5 \degree$ and $H_{\rm off} =
    -760$ Oe.}
  \label{figdisk2}
\end{figure}

The mechanical-FMR study on the $D_2=0.5~\mu$m disk was performed with
a probe-sample separation $s\approx1.0~\mu$m, kept constant at the
four different frequencies studied and shown in Fig.\ref{figdisk2}. As
for the $D_1=1.0~\mu$m disk, different magnetostatic modes can be
observed, with the most intense at high field. The main difference
with the larger disk is that this largest peak happens at much lower
field ($\approx 2.5$ kOe less). It can not be explained solely by the
larger stray field from the probe, which is closer from the sample.
Actually, most of this shift is due to finite size effects. The latter
are also responsible for the larger separation between modes on the
smaller disk \cite{Loubens07}. A detailed analysis of the spectral
features (position, amplitude and line width) of this disk is done in
section \ref{comparison}.

We note in passing that the signal to noise ratio obtained at
$T=280~K$ with a time constant of one second on this tiny sample is
very decent. One has to remember, that the angle of precession is less
than $1 \degree$. This translates in a spin sensitivity of about 1000
spins for our mechanical-FMR setup. We also recall that for this
sample, the probe is not optimal, since $\Psi/D_2=7>2$.

\section{Modeling of the spectra \label{models}}

We shall present below two different approaches to analyze the
experimental data. The first one is a 2D approximate analytical model,
which assumes an homogeneous magnetization dynamics along the
thickness. The second one is a full 3D micromagnetic simulation of the
dynamical susceptibility using a discretized representation of the
$0.5~\mu$m disk with regular cubic cells of 3.9 nm lateral size and
assuming an homogeneous magnetization dynamics inside each cell.

The motivation is to develop a comprehensive framework to anayze the
spectroscopic features observed experimentally by mechanical-FMR. The
parameters introduced in the models are the magnetic characteristics
of the material as extracted from cavity-FMR of the thin film and
introducing finite size effects (with the assumed diameters and
thickness of the disks) and a small misalignment $\theta_H \neq 0
\degree$ between the applied dc field and the normal of the disks to
fit the mechanical-FMR data.  To predict the effects of the
misalignment, we shall use a perturbation approach, assuming that the
profile of the modes is unchanged compared to the perfect
perpendicular alignment. By comparing the analytical models and the
simulations, we will evaluate the range of validity of this
approximation.

\subsection{2D analytical model \label{analmod}}

\begin{figure}
  \includegraphics[width=5cm]{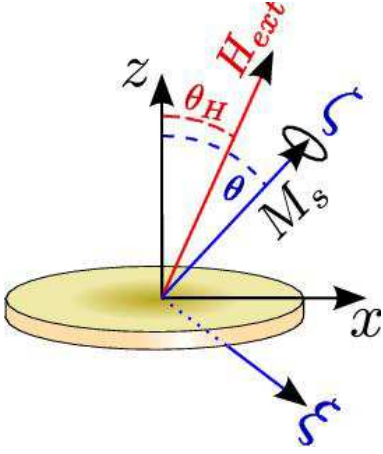}
  \caption{(Color online) Graphic representation of the Cartesian
    frame $(\xi,\zeta)$ (blue) rotated in the direction of the
    equilibrium magnetization, when the applied field makes an angle
    $\theta_H $ with the normal of the disk ($z$-axis).}
  \label{fig-zeta}
\end{figure}

In this section, we derive analytically the resonance condition that
applies to the case where the external magnetic field makes a small
angle $\theta_H$ with the normal of the disk, which breaks the axial
symmetry. The notations used below are defined in Fig.\ref{fig-zeta}.

The idea is to perform a linearization of the equation of motion for
the magnetization around the equilibrium configuration using the SW
dispersion relation in the dipole-exchange regime. We start with the
gyromagnetic term of the Landau-Lifshitz equation of motion for the
magnetization:
\begin{equation}
  \partial_t{\bm M} = \gamma\left[\bm H_{\rm eff} \times \bm M\right].
  \label{LL}
\end{equation}
The effective magnetic field
\begin{equation}
  \bm H_{\rm eff} = \bm H_{\rm ext} + \bm H_{\rm ex} + \bm H_{\rm dip}
  \label{Heff}
\end{equation}
is the sum of three fields: i) $\bm H_{\rm ext}$, the external
magnetic field, which includes the stray field produced by the probe;
ii) the exchange field
\begin{equation}
  \bm H_{\rm ex} = 4 \pi \Lambda^2 \bm\nabla^2\bm M,
  \label{Hex}
\end{equation} 
$\Lambda$ being the exchange length; iii) $\bm H_{\rm dip}$ the
internal dipolar field of the sample:
\begin{equation}
  \bm H_{\rm dip}(t, \bm r) = -4\pi\int_{r' < R} \hat{\bm G}(\bm r -
  \bm r')\cdot\bm M(t, \bm r')d^2\bm r'.
  \label{Hdip}
\end{equation}
Here $R$ is the radius of the disk, and $\hat{\bm G}(\bm r)$ is the
dipolar Green tensor, that has the form
\begin{equation}
  \hat{\bm G}(\bm r) = \int \widehat{\bm G}_{\bm k}e^{i\bm k\cdot\bm
    r} \frac{d^2\bm k}{(2\pi)^2},
  \label{Gr}
\end{equation}
which introduces non-local interactions. The Fourier image $\hat{\bm
  G}_{\bm k}$ of the dipolar Green tensor for the lowest SW branch
(with uniform magnetization across the film thickness) is given by
\cite{Kalinikos1986}
\begin{equation}
  \hat{\bm G}_{\bm k} = P_k \bm z\bm z + (1 - P_k)\frac{\bm k\bm
    k}{k^2},
  \label{Gk}
\end{equation}
where
\begin{equation}
  P_k = \frac{1 - e^{-k t}}{kt}.
  \label{Pk}
\end{equation}
$\bm z$ is the unit vector orthogonal to the disk plane, $t$ is the
thickness of the ferromagnetic film, and $\bm k$ the SW wave vector.

For a {\em thin} magnetic disk (with aspect ratio $t/R \ll 1$) we can
neglect non-uniformities of the equilibrium magnetization distribution
$\bm M(\bm r)$. It will be assumed that $\bm M(\bm r) = M_s\bm\zeta$,
where $\bm\zeta$ is a constant unit vector along the effective
magnetic field direction. At equilibrium, a uniform magnetization will
create a non-uniform dipolar field $-4\pi M_s\hat{{\bm N}}
\cdot\bm\zeta$, where $\hat{{\bm N}}$ is the position-dependent static
demagnetization tensor defined in appendix \ref{appA}. The unit vector
$\bm \zeta$ is determined by the condition:
\begin{equation}
  \left(\bm H_{\rm ext} -4\pi M_s\hat{\overline{\bm
        N}}\cdot\bm\zeta\right) \quad||\quad \bm\zeta.
  \label{eq-cond}
\end{equation}
This can be rewritten as:
\begin{equation}
  H_{\rm ext} \sin (\theta -\theta_H ) + 2 \pi M_s \overline{N_{xx}} \sin 2 \theta - 2\pi M_s \overline{N_{zz}} \sin 2 \theta = 0,
\end{equation}
allowing to determine the equilibrium angle of the magnetization with
respect to the normal of the disk. $\hat{\overline{\bm N}}$ is the
averaged demagnetization tensor over the volume of the disk:
\begin{equation}
  \hat{\overline{\bm N}} = \frac{1}{\pi R^2}\int_{r < R} \hat{\bm
    N}(\bm r)d^2\bm r.
  \label{N-ave}
\end{equation}

For small perturbations around the equilibrium state, the transverse
part of the magnetization $\bm m(t, \bm r) = \bm M(t, \bm r) -
M_s\bm\zeta$ obeys the linear equation:
\begin{eqnarray}
  \partial_t{\bm m} & = &
  \gamma\left[\left(\bm H_{\rm ext} -4\pi M_s \hat{{\bm
          N}}\cdot\bm\zeta\right) \times \bm m\right] \nonumber \\
  &	&+\gamma M_s\left[\left(\bm h_{\rm dip} + 4\pi
      \Lambda^2 \bm\nabla^2\bm m\right) \times \bm\zeta\right],
  \label{eq-m}
\end{eqnarray}
where
\begin{equation}
  \bm h_{\rm dip}(t, \bm r) = -4\pi\int_{r' < R} \hat{\bm G}(\bm
  r - \bm r')\cdot\bm m(t, \bm r')d^2\bm r'.
  \label{hdip}
\end{equation}

We recall that for axially symmetric samples with negligible
thickness, the normal modes are of the form
\begin{equation}
  \bm m_{\ell, m}(t, \bm r) = {\cal J}^{\ell}_m (\bm r) \bm\mu_{\ell,
      m}(t) = J_\ell(k_{\ell,m}r)\cos(\ell\phi) \bm\mu_{\ell,m}(t),
  \label{profile}
\end{equation}
where $\bm\mu_{\ell, m}$ is a left circularly polarized unit vector
rotating at $\omega_s$, $\ell$ and $m$ are, respectively, the
azimuthal and radial mode indices, $J_\ell(x)$ are Bessel functions,
and the wave numbers $k_{\ell,m}$ are determined by the boundary
conditions.

For a thin disk one can assume that the mode profiles remain the same
as in the limit $t \to 0$. Substituting Eq.(\ref{profile}) into
Eq.(\ref{eq-m}), multiplying by $J_\ell(k_{\ell, m}r)\cos(\ell\phi)$,
and averaging over the disk area, one obtains the usual differential
equation for the unit vector $\bm\mu_{\ell, m}$:
\begin{eqnarray}
  \partial_t{\bm\mu_{\ell, m}} & = &
  \gamma\left[\left(\bm H_{\rm ext} -4\pi M_s
      \hat{\widetilde{\bm N}}_{\ell,m} \cdot\bm\zeta\right) \times
    \bm\mu_{\ell, m}\right] \nonumber \\
  & &	-4\pi\gamma M_s\left[\left(
      \hat{\bm G}_{\ell, m}\cdot\bm\mu_{\ell, m} + (\Lambda k_{\ell, m})^2\bm\mu_{\ell, m}
    \right) \times \bm\zeta\right].
  \label{eq-mu}
\end{eqnarray}
where $\hat{\widetilde{\bm N}}_{\ell,m}$ has been defined in
Eq.(\ref{N-dyn}).

The tensor $\hat{\bm G}_{\ell, m}$ is equal to
\begin{widetext}
\begin{eqnarray}
  \hat{\bm G}_{\ell, m} &=& \frac{1}{C_{\ell, m}}\int_{r < R} 
  \int_{r' < R} \hat{\bm G}(\bm r - \bm r')J_\ell(k_{\ell, m} r)J_\ell(k_{\ell, m} r')e^{i\ell(\phi' - \phi)}
  d^2\bm r' d^2\bm r\\
  &=& G_{\ell, m}^{\perp}\bm z\bm z + (1 - G_{\ell,m}^{\perp})(\bm
  x\bm x), \nonumber
  \label{G-mode}
\end{eqnarray}
where
\begin{equation}
  G_{\ell, m}^{\perp} = \frac{2 \pi R^2}{C_{\ell, m}} \int_0^\infty
  \left(\frac
    {kJ_{\ell-1}(kR)J_\ell(k_{\ell, m}R) - k_{\ell, m}J_{\ell-1}(k_{\ell, m}R)J_\ell(kR)}
    {k^2 - k_{\ell, m}^2}\right)^2P_k k dk.
  \label{G-perp}
\end{equation}
\end{widetext}
$P_k$ has been defined in Eq.(\ref{Pk}) and the expression for the
normalization constant $C_{\ell, m}$ is given in Eq.(\ref{Clm}) and
simplifies to
\begin{equation}
  C_{\ell, m} =  \pi R^2\left( J_{\ell + 1}(k_{\ell, m} R)J_{\ell - 1}(k_{\ell, m} R)\right).
  \label{C}
\end{equation}

Eqs.(\ref{eq-mu}) is a system of ordinary differential equations with
constant coefficients. The tensors $\hat{\bm N}_{\ell, m}$ and
$\hat{\bm G}_{\ell, m}$ are given by one-dimensional integrals and are
computed numerically. The resonance field is obtained by solving
Eqs.(\ref{eq-mu}) (see appendix \ref{appB}). It yields
\begin{subequations}
  \begin{alignat}{3}
    \partial_t{m_x} & = & -\omega_1 m_y \\
    \partial_t{m_y} & = & + \omega_2 m_x
  \end{alignat}
  \label{omega-t}
\end{subequations}
where
\begin{widetext}
  \begin{subequations}
    \begin{alignat}{3}
      \frac{\omega_1}{\gamma} & = & {H}_{\rm ext}
      \cos(\theta-\theta_H)+2\pi M_s \left(
        -\widetilde{N_{zz}}-\widetilde{N_{xx}} - \left( +
          \widetilde{N_{zz}} -\widetilde{N_{xx}} \right)\cos(2\theta)
      \right) + 4\pi M_s \Lambda^2 k_{\ell,m}^2 \\
      \frac{\omega_2}{\gamma} & = & {H}_{\rm ext}
      \cos(\theta-\theta_H)+2\pi M_s \left(
        -\widetilde{N_{zz}}-\widetilde{N_{xx}} - \left( +
          \widetilde{N_{zz}} -\widetilde{N_{xx}} \right)\cos(2\theta)
      \right) + 4\pi M_s \Lambda^2 k_{\ell,m}^2 + 2 \pi M_s (1- 2
      G_{\ell, m}^{\perp}) \cos(2 \theta).
    \end{alignat}
  \label{tyb-arb-theta-1} 
\end{subequations}
\end{widetext}

This system of homogeneous equation has periodic solutions if the
determinant of the characteristic system is equal to zero. This leads
finally to the quadratic solution
\begin{equation}
  \omega_s^2= \omega_1 \omega_2,
  \label{tyb-arb-theta-2}
\end{equation}
which defines the resonance condition for this geometry.

\subsection{3D micromagnetic simulation}

\begin{figure}
  \includegraphics[width=8.5cm]{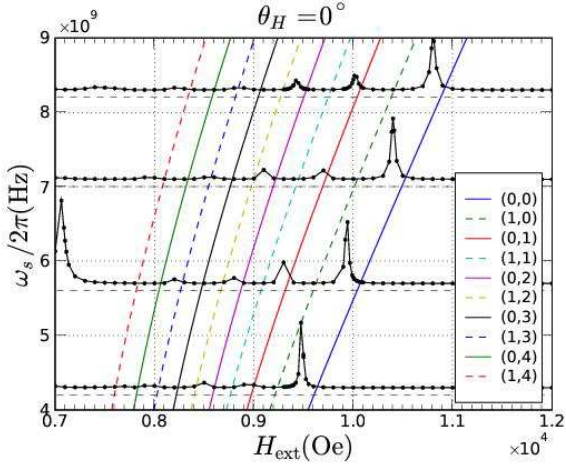}
  \caption{(Color online) Comparison between the 2D analytical model
    (oblique lines) and the 3D micromagnetic simulation (horizontal
    spectra) for the prediction of the resonance fields of a
    $D_2=0.5~\mu$m Py disk magnetized at $\theta_H =0 \degree$. The
    continuous lines are the locus of the resonance field of the
    radial modes $(\ell=0,m)$ predicted analytically (dashed lines
    correspond to the hidden modes $(\ell=1,m)$).}
  \label{simu-0d}
\end{figure}

The dynamical susceptibility spectra of the smallest Py disk has been
simulated by a micromagnetic code developed by S. Labb\'e
\cite{Labbe1999} and later by F. Boust and N. Vukadinovic
\cite{Boust2004}. In this approach, the disk volume is discretized by
a regular cubic mesh of total size $128\times128\times11$, where each
cube has an edge size of 3.9 nm. The magnetization vector is assumed
to be uniform inside each cell. This approximation is valid only
because the cell size is smaller than the exchange length $\Lambda =
5$~nm for Py. Two 3D codes are used to calculate the dynamical
response. For each value of the external field $H_{\rm ext}$, the
first code calculates the stable configuration of the magnetization
vector $\bm M(\bm r)$ by solving the Landau-Lifshitz equation in the
time domain. Hence, this code allows to incorporate in the simulation
the small spatial dependence of the direction of $\bm M$ inside the
sample volume in the quasi-saturated state. The second code computes
the full dynamic susceptibility tensor $\hat{\chi}$ from the
linearization of the Landau-Lifshitz equation around the local
equilibrium configuration.  The used material parameters are identical
to the ones measured in the extended thin film ($\gamma = 1.849~10^7$
rad.sec$^{-1}$.G$^{-1}$, $4\pi M_s=9775$ G, $A=10^{-6}$ erg/cm,
$\alpha=6~10^{-3}$).

\begin{figure}
  \includegraphics[width=7.0cm]{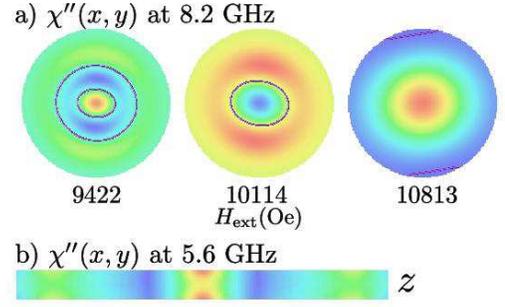}
  \caption{(Color online) a) Images of the transverse susceptibility
    obtained through the 3D micromagnetic simulation for the first
    three modes at 8.2 GHz of a $D_2=0.5~\mu$m Py disk magnetized at
    $\theta_H =0 \degree$. The thick lines indicate the nodes of
    precession.  b) Image of the transverse susceptibility for the
    third mode (at $H_{\rm ext}=8.8$ kOe) of the 5.6 GHz spectrum
    showing that the magnetization dynamics is not uniform across the
    thickness of the disk.}
  \label{simu-modes}
\end{figure}

At first, we shall compare the spectra calculated at $\theta_H=0
\degree$ by the 3D micromagnetic simulation and the 2D analytical
model.  Fig.\ref{simu-0d} shows the amplitude of the imaginary part of
the simulated susceptibility of the in-plane component $\chi"_{xx}$ at
four different frequencies: 8.2, 7.0, 5.6 and 4.2 GHz. A series of
quantized modes is observed in the 8.2 GHz spectrum.  The spatial
distribution of the resonant modes in the mid-plane of the disks are
shown in Fig.\ref{simu-modes}a for the three most intense peaks. The
observed profiles correspond to the expected eigen-modes ${\cal
  J}^\ell_m (\bm r)$, with $\ell=0$ and $m=0,1$ and 2
respectively. For comparison we have also plotted in Fig.\ref{simu-0d}
the locus of the resonances (see oblique lines) predicted by the 2D
analytical model using the same parameters as in the 3D model. The
agreement between the two models is excellent for the lowest energy
(highest field) modes.

In Fig.\ref{simu-modes}b, the calculated transverse susceptibility of
the third mode (at $H_{\rm ext}=8.8$ kOe) of the 5.6 GHz spectrum is
shown along the thickness of the disk.  It shows that the
magnetization dynamics is not uniform along the thickness,
particularly in the center of the disk. This is due to the
non-uniformity of the internal field along the thickness, as the disk
is magnetized along its normal. This effect would get stronger if the
thickness of the disk would be larger, and could eventually lead to an
edge mode. In fact, the localization of the lowest energy mode at the
top and bottom surfaces of Cu/Py/Cu sub-micron size disks, where the
thickness of the Py layer was 100 nm, was observed experimentally and
calculated by full 3D simulations of $\chi"$ \cite{Loubens07}.

\begin{figure}
  \includegraphics[width=8.5cm]{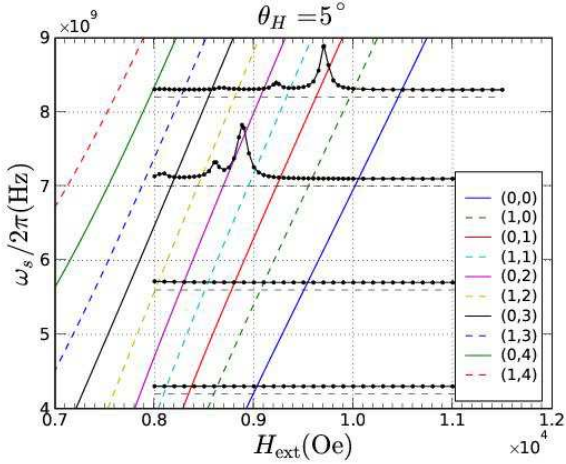}
  \caption{(Color online) Comparison between the 2D analytical model
    (oblique lines) and the 3D micromagnetic simulation (horizontal
    spectra) for the prediction of the resonance fields of a
    $D_2=0.5~\mu$m Py disk magnetized at $\theta_H =5 \degree$. The
    continuous lines are the locus of the resonance field of the
    radial modes $(\ell=0,m)$ predicted analytically (dashed lines
    correspond to the hidden modes $(\ell=1,m)$).}
  \label{simu-5d}
\end{figure}

\begin{figure}
  \includegraphics[width=8.5cm]{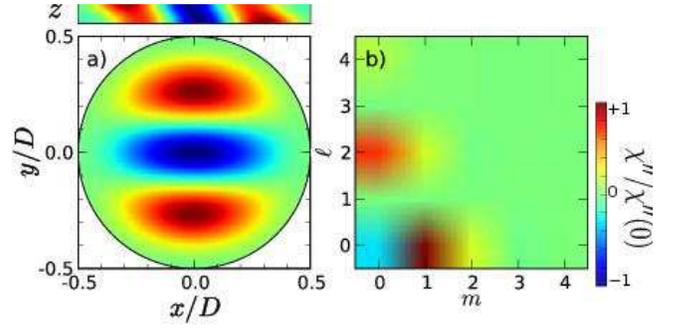}
  \caption{(Color online) a) Image of the spatial distribution of the
    resonant profile for the 2nd peak at 9.2 kOe on the 8.2 GHz
    simulated spectrum of Fig.\ref{simu-5d}. The cartography is shown
    both in the axial and radial middle sections. b) The mode shown in
    a) is now decomposed in the ${\cal J}^\ell_m$ basis. The figure reveals
    here the spectral weights of the different eigen-modes.}
  \label{new-mode}
\end{figure}

We have also performed micromagnetic simulations of the dynamic
susceptibility at $\theta_H \neq 0^\circ$, using the same value for
the parameters as above. Fig.\ref{simu-5d} shows the simulated
spectrum at $\theta_H=5^\circ$ for the four same excitation
frequencies. The simulations are limited to the quasi-saturated field
range when $H_{\rm ext} > 8$ kOe. The highest frequency spectrum at
8.2 GHz shows two resonances. The spatial profile (not shown) of the
lowest energy mode (located at 9.7 kOe on the 8.2 GHz spectrum)
resembles ${\cal J}^0_0(\bm r)$ described above. Fig.\ref{new-mode}a
shows the spatial profile for the second peak, at 9.2 kOe on the 8.2
GHz spectrum. The observed spatial profile breaks the axial symmetry
due to the small tilt angle that the magnetization makes with the disk
normal. This profile can be decomposed in the basis of the normal
modes ${\cal J}^\ell_m$.  The results is shown in Fig.\ref{new-mode}b,
where we find that
\begin{eqnarray} 
  \chi"(\bm r) &\approx& -\frac{3}{4} \left \{  \frac{1}{3} {\cal J}^0_0(\bm r) - {\cal J}^0_1(\bm r)   - \frac{2}{3} {\cal J}^2_0(\bm r) \right \}  \nonumber \\ 
  & & +\frac{1}{5} \left \{  {\cal J}^0_2(\bm r) + {\cal J}^2_1(\bm r) + {\cal J}^4_0 (\bm r) \right \}.
\end{eqnarray}

Hence, the normal modes in the presence of a small tilt angle
$\theta_H$ are no longer the functions ${\cal J}^\ell_m $ but rather a
linear combination of them. We have also plotted in Fig.\ref{simu-5d}
the 2D analytical prediction of the resonance fields as oblique
lines. We find an important discrepancy between the two models in the
location of the resonance for the fundamental mode (almost 850 G
apart), but also in the effective gyromagnetic ratio. In order to fit
the 3D simulated spectra with the 2D analytical model and $\theta_H$
as the adjustable parameter, a much larger value of $\theta_H \approx
10\degree$ has to be used in the analytical model than in the
simulation. This shows the limits of the two main approximations of
the 2D model: homogeneous dynamics along the disk thickness and
unperturbed ${\cal J}^\ell_m $ normal modes basis to compute the
resonance fields.

\section{Comparison with the experimental data \label{comparison}}

In this section, we want to analyze the three spectroscopic
informations that are i) the position ii) the amplitude and iii) the
width of the resonance peaks measured experimentally on the Py disks
of thickness 43.3 nm and of diameters $D_1=1.0~\mu$m and
$D_2=0.5~\mu$m.

\subsection{Position}

\begin{figure}
  \includegraphics[width=8.5cm]{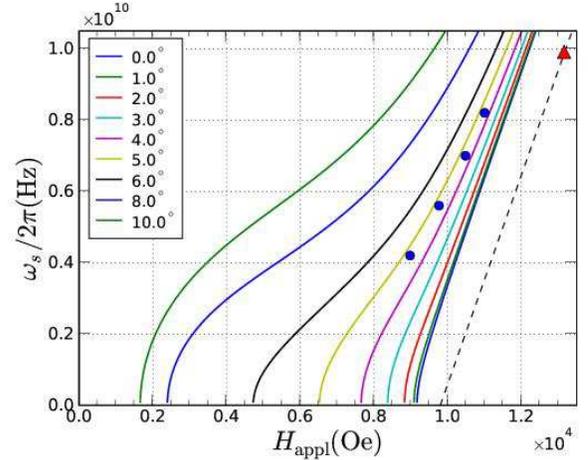}
  \caption{(Color online) Resonance field (blue dots) of the
    fundamental mode measured on the $D_1=1.0~\mu$m disk along with
    the analytically predicted angle dependence of
    $H_\text{res}(\omega_s)$ for different values of $\theta_H$
    (lines). Also indicated is the measured resonance position in the
    extended film (red triangle) and the gyromagnetic ratio (dashed
    line): the shift with the $\theta_H = 0^\circ$ line is due to
    finite size effects in the disk.}
  \label{figpos}
\end{figure}

As mentioned earlier, the values of the resonance fields are an
important indicator of the effective internal field, which should not
be disregarded by using the magnetization as a fitting parameter. We
shall rely on the value of $M_s$ measured by cavity-FMR on the
extended thin film used for the fabrication of the disks and we shall
assume that $4\pi M_s=9775$ G is unchanged after the patterning
process. The cavity-FMR also provides the value of $\gamma =
1.849~10^7$ rad.sec$^{-1}$.G$^{-1}$ for our alloy composition. We
first concentrate on the position of the fundamental mode (highest
field mode) measured on the largest disk, of diameter $D_1=1.0~\mu$m
(see Fig.\ref{figdisk1}). The blue dots shown in Fig.\ref{figpos} are
the positions of the main peak put in a $(\omega_s,H_{\rm appl})$
diagram. Also on the figure (red triangle) is the resonance position
of the uniform mode measured by the cavity-FMR, and the dashed line is
the extrapolated locus of the mode as a function of frequency. We use
the 2D analytical model presented in section \ref{analmod} to fit the
resonance fields. The results are shown in continuous lines. The two
fitting parameters are the angle $\theta_H$ and the offset field
induced by the tip $H_{\rm off}$.  Changes of $H_{\rm off}$
corresponds simply to an over-all shift of the whole set of curves. At
$H_{\rm off}=0$ Oe, the $\theta_H=0^\circ$ line is already shifted by
almost 1 kOe lower in field (higher in energy) compared to the
extended film. This is due to the cost in exchange and dipolar
energies in the nanostructure. Increasing $\theta_H$ in
Fig.\ref{figpos} amplifies this shift but it also reduces the
effective gyromagnetic ratio $\gamma_{\rm eff}$, which becomes
frequency dependent. There is a unique value of $\theta_H$ that
reproduces the experimentally observed frequency dependence of
$\gamma_{\rm eff}(H)$ and the best fit is obtained for $\theta_H
=4.9\degree$. Such a value for $\theta_H$ is plausible, since our
setup does not incorporate a precise goniometer, and the angular
orientation is difficult to tune on small size samples. The fit also
provides a value for $H_{\rm off}=-70$ G.  This value is close to the
expected shift induced by the stray field of the tip at the distance
$s=3.1~\mu$m of the disk.

Using the fit values above, we have reported on Fig.\ref{figdisk1} the
locus of the higher order modes. The continuous lines indicate the
cylindrically symmetric modes ($\ell=0$) and the dashed lines
indicates the modes with non-zero angular numbers $\ell\neq 0$, which
should be hidden due to the homogeneous excitation field $h$. We find
a good agreement with the data, except at the lowest frequency (4.2
GHz) where a peak appears between the ${\cal J}^0_0$ and ${\cal
  J}^0_1$ modes. This new resonance is actually close to the location
of the ${\cal J}^{1}_0$ hidden mode. A possible explanation is that
this resonance is a reminiscence of the broken symmetry mode found in
the simulation of the $D_2 = 0.5$ $\mu$m disk at $\theta_H = 5
\degree$. Such a mode, being a combination of mainly ${\cal J}^{0}_0$
and ${\cal J}^{0}_1$, would resonate between the $m=0$ and $m=1$
mode. This mode becomes prominent at the lowest frequency because, as
the excitation frequency decreases, the applied field becomes lower
and the magnetic configuration becomes more sensitive to the in-plane
component of the applied field, which favors the excitation of a mode
having the symmetry of the in-plane component.

The differences between the data and the 2D analytical model can be
due to the approximations and assumptions made in the model. One
factor that affects the peak position is the value of the pinning
condition. The expressions derived above in the 2D approximate model
assume a total pinning of the SW at the periphery. Assuming no pinning
at all at the periphery would move the peak position upward in field
by about 50 G. It corresponds to keeping the total pinning condition
and increasing the disk diameter by some amount of the order of the
film thickness \cite{Guslienko2002}. Also, the radial component of the
stray field of the probe is not taken into account in the model. This
would accentuate the effects of $\theta_H$ on the static configuration
of $\bm M(\bm r)$ in the sample. Other experimental uncertainties
would be sufficient to account for the small discrepancy found in our
analysis. We recall that the diameter of the disk used in the model is
the one measured by SEM (see Fig.\ref{disks}). A peripheral oxidation
of the alloy can not be excluded, since there is no protective alumina
on the periphery of the disks. Change of the disk diameter in the
analytical model would shift the resonance peaks and affect the field
separations between them. Also, the separation between the probe and
the sample, and the magnetic moment of the spherical probe, are known
within 10\%.  This shows the limits of using the mechanical-FMR for a
precise determination of the unperturbed FMR peak positions of a
sub-micron size sample.

Finally we have repeated this analysis on the smaller disk, with a
diameter of $D_2=0.5~\mu$m. Fitting the data with the analytical model
gives $\theta_H = 9.5 \degree$ and $H_{\rm off} = -760$ Oe. We have
reported with straight lines on Fig.\ref{figdisk2} the results of the
analytical model. The value of $\theta_H$ found here is in
disagreement with the previous finding, which is somewhat surprising
since the two disks are located nearby, on the same substrate. Several
arguments suggest that the true value of $\theta_H$ is the value for
the largest disk, \emph{i.e.} $\approx 5 \degree$. As shown in
Fig.\ref{simu-5d}, the analytical model tends to underestimate the
effect of the angle as the disk shrinks in diameter, compared to the
3D simulation. This affects the shift in the resonance field and the
lowering of $\gamma_{\rm eff}$, both underestimated by the 2D
approximate model. A fit with this model will thus yield to a larger
$\theta_H$ value than the 3D simulation.  If we now compare the
experimental results of Fig.\ref{figdisk2} with the 3D simulation at
$\theta_H =5 \degree$ of Fig.\ref{simu-5d}, we find that $H_{\rm off}
\approx -1$ kOe, which is a shift larger than the one expected,
$\approx -520$ Oe from $s=1.0~\mu$m (see section \ref{shift}). We
emphasize here that the uncertainties mentioned above in the
properties of the sample and the probe could translate in substantial
errors in the 2D and 3D analytical models. In fact, finite size
effects in the $D_2=0.5~\mu$m crucially depend on the exact properties
of the sample. Moreover, due to the small separation between the probe
and the sample, the effect of the radial stray field from the probe is
also more important than for the largest disk.

\subsection{Amplitude}

\begin{figure}
  \includegraphics[width=8.5cm]{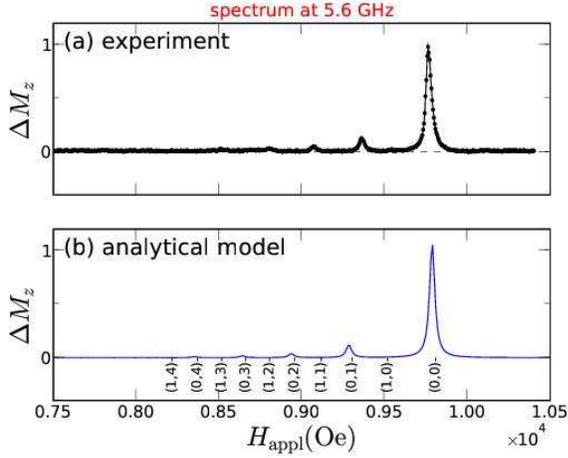}
  \caption{(Color online) Comparison between experiment and analytical
    model of the amplitude of the peaks for the disk of diameter 1.0
    $\mu$m at 5.6 GHz.}
  \label{figamp}
\end{figure}

The amplitude of the modes indicates their coupling with the
spectrometer. In the following we attempt to predict the relative
amplitudes between the peaks. We will concentrate on the 5.6 GHz
spectrum measured for the largest disk (cf. Fig.\ref{figdisk1}).

Since the mechanical-FMR is using different schemes for the excitation
and detection part, they need to be treated separately. We first
compute the coupling to the excitation, which is identical to all FMR
spectrometers. The microwave field being uniform at the scale of the
sample, the coupling is simply given by the overlap integral with the
transverse susceptibility. In our notation, we find that the angle of
precession is given by
\begin{equation}
  \vartheta_m = \frac{2 h}{\alpha M_s R^2} \int_{r<R} {\cal J}^0_m (r) r d{r} 
\end{equation}
where we only consider the coupling to the $(\ell=0,m)$ modes and
where $\alpha$ is the damping coefficient.

In our case, the force induced on the cantilever is given by
\begin{equation}
  F_{z,m} = m_{\rm sph} \int_{V_s} \Delta M_{z,m} ({\bm r}) {\EuScript
    G}_{zz}(r,z+s+\Psi/2) d^2r dz,
  \label{force}
\end{equation}
where the integral is the gradient of field along the $z$ direction
induced by the local variation of longitudinal component of
magnetization inside the sample: $\Delta M_{z,m}({\bm r}) = \frac 12
M_s \vartheta_m^2 {\cal J}^\ell_m({\bm r})^2/C_{\ell,m}$ and where
\begin{equation}
 {\EuScript G}_{zz} (r,z) = \frac{9 z}{(r^2+z^2)^{5/2}} -\frac{15 z^3}{(r^2+z^2)^{7/2}}.
\end{equation}
In Eq.\ref{force}, we have used the fact that our probe has a
spherical shape and can be viewed as a magnetic dipole $m_{\rm sph}$
placed at its center.

We have reported in Fig.\ref{figamp} a comparison between the measured
spectra and the calculated peak shape using the analytical model and
assuming that the line width is identical for all the modes. We find
that the agreement with the data is excellent. In particular, we
observe that the peak amplitude decreases by almost an order of
magnitude between the ${\cal J}^{0}_0$ and ${\cal J}^{0}_1$
modes. This decrease is less pronounced in conventional FMR which
measures the transverse component of the magnetization (see
Fig.\ref{simu-0d}). This implies some caution when comparing the
relative amplitude of the peaks found in the 3D simulation of the
transverse susceptibility with the one found in the mechanical-FMR
experiment.

\subsection{Line width}

\begin{figure}
  \includegraphics[width=8.5cm]{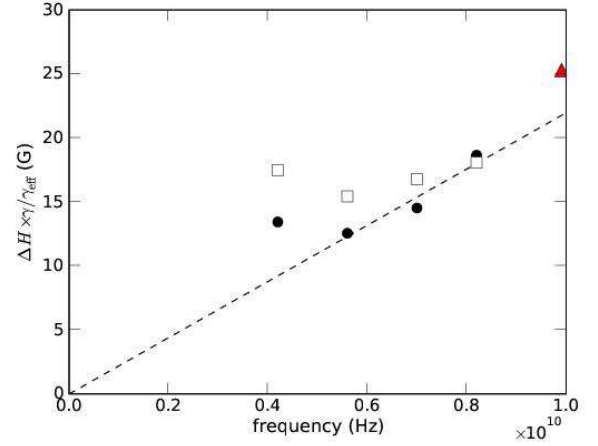}
  \caption{(Color online) Frequency dependence of the line width
    measured on both the $D_1=1.0~\mu$m disk (full circles) and the
    $D_2=0.5~\mu$m disk (open boxes). The red triangle indicates the
    value measured by cavity-FMR on the extended film on Si. The
    mechanical-FMR data are renormalized by $\gamma_{\rm eff}$. The
    dashed line corresponds to a damping coefficient $\alpha = 6
    \times 10^{-3}$.}
  \label{figwidth}
\end{figure}

Another important characterization concerns the width $\Delta H$ of
the resonance line. The full circles and open boxes in
Fig.\ref{figwidth} are the line widths measured as a function of
frequency in the Py disks of diameters $D_1=1.0~\mu$m and
$D_2=0.5~\mu$m, respectively. All values have been renormalized by the
effective gyromagnetic ratio $\gamma_{\rm eff} = \partial \omega_s/
\partial H_{\rm appl}$. These line widths are among the smallest
reported for Py, confirming the excellent quality of the material used
for this study.

The resonance line width (half-width at mid-height of the absorption
curve) of the Py polycrystalline magnetic layer is described
\cite{Hurdequint2002} as the sum of two contributions:
\begin{equation}
  \Delta H = \frac{1}{2\gamma T_1} + \Delta H^\text{res} (\Delta H_{u},
  \Delta \theta_H).
  \label{deltaH}
\end{equation}
The first term relates to the intrinsic relaxation rate of the
magnetization vector whereas the second term corresponds to the
inhomogeneous broadening induced by a distribution of internal field,
as well as by a distribution of polarization angles, of
characteristics widths $\Delta H_{u}$ and $\Delta \theta_H$,
respectively. $\Delta H_{u}$ reflects the spatial inhomogeneity of the
field parameter $H_u$ whereas $\Delta \theta_H$ refers to the
distribution of the orientation $\theta_H$ of the applied field with
respect to the normal to the surface of the individual crystallites.

The results obtained on the disks should be compared to the value
observed by cavity-FMR on the extended thin film. In the cavity-FMR
studies, the line widths measured on the film grown on the atomically
flat Si substrate in the parallel and in the perpendicular geometries
are equal (25 Oe) and reflect almost entirely the homogeneous
contribution. The measured value is represented by a red triangle in
Fig.\ref{figwidth}.
It allows for an estimate of the intrinsic damping parameter
($\alpha=1/(2\omega T_1)$), found to be $\alpha=7~10^{-3}$. This value
is an upper bound for $\alpha$, since the presence of a small
inhomogeneous contribution in the line width would reduce the value of
the homogeneous part. The film grown on mica presents more
inhomogeneities than on Si, the results of which corresponds to a line
width increase of 3 Oe.  This is due to the even surface of the mica
substrate, which bends over because of the strain between the
different sheets of mica.
 
From the frequency dependence of the line width of the large disk, we
can estimate the damping coefficient in the nanostructure, found to be
$\alpha=6~10^{-3}$ (dashed line on Fig.\ref{figwidth}), in good
agreement with the upper bound found on the thin film. We find that
there is no inhomogeneous broadening in the nanostructure, which means
that the small amount observed in the extended film deposited on mica
is not relevant in structures confined at the sub-micron length scale
\cite{Loubens07, Chen08}. It also implies that there is no additional
broadening induced by the mechanical-FMR. Finally, the frequency
dependence of the line width at low frequencies and of the small disk
is not linear. In fact, the magnetization configuration becomes not
uniform at low applied field, and the renormalization of the field
line widths by $\gamma_{\rm eff}$ is not sufficient to recover the
intrinsic behavior. Moreover it was shown on the extended thin film
that as the applied field is decreased, the increase of the angle
between the equilibrium magnetization and the normal of the film
increases the line width, through the inhomogeneous contribution
associated \cite{Hurdequint2002} to the distribution $\Delta \theta_H$
in Eq.\ref{deltaH}.

\section{Conclusion \label{conclusion}}

It was shown in this article how MRFM can be used to detect and
\emph{quantitatively} analyze the intrinsic FMR spectra of
\emph{individual} sub-micron size samples. To realize this objective,
the coupling between the magnetic probe attached to the cantilever and
the sample has to be optimized. On one hand, it has to be as strong as
possible to detect FMR in tiny samples. On the other hand, the
inhomogeneity induced by the stray field of the probe has to be small
compared to the internal dipolar field inhomogeneity in the sample, in
order to detect the intrinsic behavior of the latter. Using a magnetic
sphere whose size is of the order of the disk-shaped samples enables
to meet these two requirements. We demonstrate a 1000 spins
sensitivity at room temperature. Using two approximate 2D and 3D
models, we can understand the measured FMR-spectra on Py disks
patterned out from the same extended film, whose characteristics are
well known. It requires a good understanding of finite size effects
and of the homogeneous shift of the the SW modes resonance fields
induced by the probe.

Finally, we would like to summarize the main advantages of
mechanical-FMR: i) its sensitivity to detect a single magnon
excitation in a buried hybrid structure, \emph{e.g.} below contact
electrodes, ii) its versatility as a near field technique (\emph{i.e.}
only sensitive to the area directly underneath the probe), which
allows spatial imaging of the magnetization dynamics, and iii) its
ability to measure the longitudinal component of the magnetization, a
quantity directly linked to the damping.

We are greatly indebted to O. Acher and A.-L. Adenot for their help
and support. This research was partially supported by the ANR
PNANO06-0235 and by the European Grant NMP-FP7 212257-2 MASTER.

\appendix

\section{Depolarization factors of a cylinder} \label{appA}

\setcounter{equation}{0}

The analytical formula that have been used for the demagnetization
tensor, $\hat{\bm N}$, of a disk of radius $R$ and thickness $t$ are
explicitly written in this appendix. The formula are actually derived
from the published work of S. Tandon \cite{Tandon2004b} but we have
chosen to reprint them below because a couple of small typos remain in
the original paper. The only assumption made here is that the
magnetization is homogeneous inside the cylindrical volume.

Because of the axial symmetry, the values of the tensor $\hat{\bm N}$
are better expressed in the cylindrical coordinates $(r,z)$ and, using
the notation of Ref.\cite{Tandon2004b}, we introduce the reduced units
$\zeta = z/R, \tau = t/(2R), \rho = r/R$:
\begin{subequations}
  \begin{alignat}{2}
    N_{zz}(r,z) = & + \frac{1}{2} \left\{ s_{\zeta,\tau} I_0(\rho,\alpha_{-}) + I_0(\rho,\alpha_{+}) \right\} \\
    N_{zr}(r,z) = & - \frac{1}{2} \left\{ I_1(\rho,\alpha_{-}) - I_1(\rho,\alpha_{+}) \right\} \\
    N_{rr}(r,z) = & + \frac{1}{4}\left\{ s_{\zeta,\tau} I_2
      (\rho,\alpha_{-}) + I_2(\rho,\alpha_{+}) - 2 H_{\tau,\zeta}
      I_2(\rho,0) \right\} \nonumber \\
    & - \frac{1}{4}\left\{ s_{\zeta,\tau} I_0 (\rho,\alpha_{-}) +
      I_0(\rho,\alpha_{+}) - 2 H_{\tau,\zeta} I_0(\rho,0) \right\}
  \end{alignat}
\end{subequations}
where the notations $\alpha_{-} = | \zeta-\tau | $ and $\alpha_{+}= |
\zeta + \tau |$ are respectively the distance (in reduced units) with
the bottom and top surface of the cylinder. The function $s$ and $H$
design respectively the Sign and Heavisde functions:
\begin{eqnarray}
  s_{x,y} & = & \begin{cases} +1 \text{\hspace{2mm} if \hspace{2mm}} x < y \nonumber\\ -1
    \text{\hspace{2mm} else \hspace{2mm}} \end{cases} \nonumber\\
  H_{x,y} & = & \begin{cases} 1 \text{\hspace{2mm} if \hspace{2mm}} x > y \nonumber\\ 0
    \text{\hspace{2mm} else \hspace{2mm}} \end{cases} \nonumber
\end{eqnarray}

The integrals $I_i$ have the following expressions: 
\begin{subequations}
  \begin{alignat}{2}
    I_0 (\rho,\alpha) = & s_{1,\rho} \frac 12 \Lambda_0(\beta,\kappa)-
    \frac{k \alpha}{2 \pi \sqrt{\rho}} K(k) + H_{1,\rho}\\
    I_1 (\rho,\alpha) = & \frac{1}{\pi k \sqrt{\rho}} \left\{ (2-m) K(k) -2 E(k) \right\}\\
    I_2 (\rho,\alpha) = & \frac{2 \alpha}{\pi k \rho^{3/2}} E(k) -
    (\alpha^2+\rho^2+2)
    \frac{\alpha k}{2 \pi \rho^{5/2}} K(k) \nonumber\\
    & -s_{1,\rho} \frac{1}{2\rho^2}
    \Lambda_0(\beta,\kappa)+\frac{H_{1,\rho}}{\rho^2}
  \end{alignat}
  \label{integ}
\end{subequations}
where
\begin{subequations}
  \begin{alignat}{3}
    m & = & k^2 = \sin^2 \kappa = \frac{4 \rho}{(\rho+1)^2+\alpha^2},\\
    \beta & = & \text{arcsin} \left(\frac{\alpha}{\sqrt{(\rho-1)^2+\alpha}} \right).
  \end{alignat}
\end{subequations}
$K(k)$ and $E(k)$ are the complete elliptic integrals of the first and
second kind and $\Lambda_0$ is the Heuman's Lambda function.

The above expressions are valid everywhere in space. The magnetic
field induction at every point in space (inside or outside the
sample's volume) simply obeys the formula:
\begin{equation}
  B_z(r,z) = H_{\rm ext} + 4 \pi M_s \left \{ \Theta(r,z)-N_{zz}(r,z)
  \right \},
\end{equation}
where $\Theta$ is a function equals to 1 inside the cylindrical volume
and 0 otherwise:
\begin{eqnarray}
  \Theta(r,z)_{x,y} & = & \begin{cases} 1 \text{\hspace{2mm} if
      \hspace{2mm}} r < R \text{ and } \left | z \right | < t/2 \nonumber\\ 0
    \text{\hspace{2mm} else. \hspace{2mm}} \end{cases} \nonumber
\end{eqnarray}

\section{Linearization in the local frame} \label{appB}

The notations have been defined in Fig.\ref{fig-zeta}: $(x,y,z)$ is
the Cartesian frame along the principal axis of the disk, with $z$
oriented along the normal and $(\xi,y,\zeta)$ the Cartesian frame of
the magnetization dynamics, with $\zeta$ along the equilibrium (or
effective magnetic field) direction. Both frames are related by a
rotation of an angle $\theta$ around the $y$ direction.  If $\theta_H$
is the angle that the external magnetic field makes with the normal of
the disk, then $\theta$ is implicitly defined by the equilibrium
condition:
\begin{equation}
  H_{\rm ext} \sin (\theta -\theta_H ) + 2 \pi M_s \overline{N_{xx}} \sin 2 \theta - 2\pi M_s \overline{N_{zz}} \sin 2 \theta = 0
\end{equation}

In all the expressions above, the tensors are expressed in the
Cartesian frame of the disk $(x,y,z)$. This applies for the
demagnetizing factors $\hat{\bm N}$ in appendix \ref{appA}, but also
for the matrices:
\begin{equation}
  \bm z \bm z = \begin{bmatrix} 0 & 0 & 0 \\ 0 & 0 & 0\\ 0 & 0 & 1 \end{bmatrix}
\end{equation}
\begin{equation}
  \bm x \bm x= \begin{bmatrix} 1 & 0 & 0 \\ 0 & 0 & 0\\ 0 & 0 & 0 \end{bmatrix}
\end{equation}

Solving Eq.(\ref{eq-mu}) requires to write down the different tensors
$\hat{\bm T} (=\hat{\bm N},\hat{\bm G})$ in the local frame of the
magnetization. This is achieved through the transformation
$\hat{\EuScript{R}}_\theta^t \hat{\bm T}\hat{\EuScript{R}}_\theta$, where
$\hat{\EuScript{R}}_\theta$ is the rotation matrix between the
$(x,y,z)$ and $(\xi,y,\zeta)$:
\begin{equation}
  \hat{\EuScript{R}}_\theta = \begin{bmatrix} \cos \theta & 0 & -\sin \theta \\ 0
    & 1 & 0\\ \sin \theta & 0 & \cos \theta \end{bmatrix}
\end{equation}
and the subscript $t$ stands for the transpose.

This allows us to find the expression Eqs.(\ref{tyb-arb-theta-1}),
where we have used symmetry arguments to impose
$\overline{N_{xy}}=\overline{N_{xz}}=0$.


\newcommand{\noopsort}[1]{}

\end{document}